\documentclass[journal, twoside]{IEEEtran}
\usepackage{makeidx}
\usepackage{amsmath}
\usepackage{amsfonts}
\usepackage{amssymb}
\usepackage{amsthm}
\usepackage{graphicx}
\usepackage{url}
\usepackage{algorithm}
\usepackage{algorithmic}
\usepackage[caption=false,font=footnotesize]{subfig}

\usepackage{cite}

\usepackage{color}

\newtheorem{definition}{Definition}
\newtheorem{theorem}{Theorem}
\newtheorem{corollary}{Corollary}
\newtheorem{lemma}{Lemma}

\ifCLASSINFOpdf
\else
\fi
\begin{document}
\title{Homology-based Distributed Coverage Hole Detection in Wireless Sensor Networks}

\author{Feng Yan,~\IEEEmembership{ Member,~IEEE,} Ana\"is Vergne, 
        Philippe Martins,~\IEEEmembership{Senior Member,~IEEE,}
        Laurent Decreusefond% <-this % stops a space
\thanks{F. Yan was with the Network and Computer Science Department, 
TELECOM ParisTech, Paris, France. He is currently with the Networks, Security and 
Multimedia Department, TELECOM Bretagne, Rennes, France (e-mail: feng.yan@telecom-bretagne.eu).

A. Vergne is  with the Geometrica team, Inria Saclay - Ile de France,
Palaiseau, France (e-mail: anais.vergne@inria.fr)

P. Martins and L. Decreusefond are with the Network 
and Computer Science Department, TELECOM ParisTech, Paris, France (e-mail: martins@telecom-paristech.fr, decreuse@telecom-paristech.fr.)

A part of this paper has been published in IEEE ICC 2012.}}% <-this % stops a space
%\thanks{Manuscript received April 19, 2005; revised December 27, 2012.}}


\markboth{IEEE/ACM TRANSACTIONS ON NETWORKING}%
{YAN \MakeLowercase{\textit{et al.}}: Homology-based Distributed Coverage Hole Detection in Wireless Sensor Networks}



\maketitle

\begin{abstract}
Homology theory provides new and powerful solutions to address the coverage 
problems in wireless sensor networks (WSNs). They are based on algebraic 
objects, such as $\check{\textrm{C}}$ech complex and Rips complex. 
$\check{\textrm{C}}$ech complex gives accurate information about coverage 
quality but requires a precise knowledge of the relative locations of nodes. 
This assumption is rather strong and hard to implement in practical deployments. 
Rips complex provides an approximation of $\check{\textrm{C}}$ech complex. 
It is easier to build and  does not require any knowledge of nodes location. 
This simplicity is at the expense of accuracy. Rips complex can not always 
detect all coverage holes. It is then necessary to evaluate its accuracy. 
This work proposes to use the proportion of the area of undiscovered coverage 
holes as performance criteria. Investigations show that it depends on 
the ratio between communication and sensing radii of a sensor. Closed-form 
expressions for lower and upper bounds of the accuracy are also derived. 
For those coverage holes which can be discovered by Rips complex, a 
homology-based distributed algorithm is proposed to detect them. 
Simulation results are consistent with the proposed analytical lower bound, 
with a maximum difference of 0.5\%. Upper bound performance depends on the 
ratio of communication and sensing radii. Simulations also show that the 
algorithm can localize about 99\% coverage holes in about 99\% cases.

\end{abstract}


% \fi
\IEEEpeerreviewmaketitle

\section{Introduction}
\IEEEPARstart{W}{ireless} sensor networks (WSNs) have attracted a great deal of
research attention due to their wide potential applications such as
battlefield surveillance, environmental monitoring and intrusion
detection. Many of these applications require a reliable detection of
specified events. Such requirement can be guaranteed only if the
target field monitored by a WSN contains no coverage holes, that is to
say regions of the domain not monitored by any sensor. Coverage holes
can be formed for many reasons, such as random deployment, energy
depletion or destruction of sensors. Consequently, it is essential to
detect and localize coverage holes in order to ensure the full
operability of a WSN.

There is already an extensive literature about the coverage problems in
WSNs. Several approaches are based on computational geometry with
tools such as Voronoi diagram and Delaunay triangulations, to discover
coverage holes \cite{FGG04, WCL04, ZZF09}. These methods require
precise information about sensor locations. This substantially limits
their applicability since acquiring accurate location information is
either expensive or impractical in many settings. Some other
approaches attempt to discover coverage holes by using only relative
distances between neighbouring sensors \cite{ZZF06, B08, B12}. Similarly,
obtaining precise range between neighbouring nodes is costly.

More recently, homology is utilized in
\cite{DSG05,DSG07,GM05} to address the coverage problems in
WSNs. Ghrist and his collaborators introduced a combinatorial object,
$\check{\textrm{C}}$ech complex, which fully characterizes
coverage properties of a WSN (existence and locations of holes).
Unfortunately, this object is very difficult to construct even if
the precise location information of sensors is provided.
Thus, they introduced a more easily computable complex,
Rips complex. This complex is constructed with the sole
knowledge of the connectivity graph of the network and gives an
approximate coverage by simple algebraic calculations. As regards
implementation in real WSN, these homology based methods are
necessarily centralized, which makes them impractical in large scale
sensor networks. Some algorithms have been proposed to implement the
above mentioned ideas in a distributed context, see \cite{ME06, TJ10}. 
But there are two disadvantages of these algorithms. On one hand,
these homology based algorithms are all dependent on the assumption 
that the communication radius of a sensor is smaller than $\sqrt{3}$ 
times the sensing radius of the sensor. When such assumption is not
satisfied, it is possible that Rips complex may miss some special 
coverage holes (defined as triangular holes in Section III).
In order to assess the accuracy of Rips complex based coverage hole detection,
it is thus of paramount importance to determine the proportion of the area of missed 
coverage holes. To the best of our knowledge, we are the first to investigate this problem. 
On the other hand, these algorithms try to verify coverage or detect coverage holes
by homology computation, which is generally of high complexity especially for large scale
networks. So it is necessary to
design an efficient algorithm to detect coverage holes.

The main contributions of our paper are as follows. First, we present the relationships
between $\check{\textrm{C}}$ech complex and Rips complex in terms of coverage
hole under different ratios between communication and 
sensing radii of a sensor. We find that when the communication radius is
at least two times sensing radius, if there is a hole in Rips complex,
there must be a hole in $\check{\textrm{C}}$ech complex. A hole in a
$\check{\textrm{C}}$ech complex missed by a Rips complex must be bounded
by a triangle. Based on that, a formal definition of triangular and non-triangular 
hole is presented. 

Second, for triangular holes, we derive the closed-form
expressions for lower and upper bounds of the proportion of their area under a homogeneous setting.
Such proportion is related to the ratio between 
communication and sensing radii of each sensor and three different ratios
between communication and sensing radii are investigated.

Third, for non-triangular holes, an efficient homology based distributed algorithm is proposed to detect 
them. In the algorithm, the Rips complex is first
constructed for a given WSN. Then some vertices and edges
are deleted without changing the number of holes in the original
Rips complex. After that, the edges lying on the boundary of holes
will be detected. Then coarse boundary cycles can be discovered.
Finally all boundaries of the non-triangular holes are found by
minimizing the length of coarse boundary cycles.

The remainder of the paper is organized as follows. Section II 
presents the related work. In Section III, the network model
and the formal definition of triangular and non-triangular hole
are given. Upper and lower bounds on the proportion of the area of 
triangular holes under different ratios between communication and
sensing radii are computed in Section IV. Section V 
describes the homology based distributed algorithm
for non-triangular holes detection. In Section VI, performance evaluation
of the bounds and the algorithm is given. Finally, Section VII 
concludes the paper.

\section{Related Work}
%\IEEEpubidadjcol

Since this paper aims to evaluate the ratio of the area of coverage holes missed by homology based approaches and to design coverage hole detection algorithms, we present the related work in two aspects: analytical coverage ratio evaluation
and coverage hole detection approaches. 

\subsection{Analytical coverage ratio evaluation}

Extensive research has been done to analyze coverage ratio of a WSN.
In \cite{LT04}, the authors studied the coverage properties of large-scale
sensor networks and obtained the fraction of the area covered by sensors. 
The sensors are assumed to have the same sensing range and 
are distributed according to a homogeneous Poisson point process (PPP) in plane.
In \cite{WY06}, the authors studied how the 
probability of $k$-coverage changes with the sensing radius 
or the number of sensors, given that sensors are deployed
as either a PPP or a uniform point process. 
In \cite{LR06}, the coverage problem in planar heterogeneous sensor 
networks are investigated and analytical expressions of coverage
are derived. Their formulation is more general in the sense that
sensor can be deployed according to an arbitrary stochastic distribution,
or can have different sensing capabilities or can have arbitrary
sensing shapes. In \cite{LHZ12}, a point in a plane
is defined to be tri-covered if it lies inside a triangle formed
by three nodes, and the probability of tri-coverage is analyzed.
None of them considered triangular holes, we provided some 
initial results about the ratio of the area of triangular holes in \cite{YMD12}
and further improve them in this paper.

\subsection{Coverage hole detection approaches}

Coverage hole detection approaches can be generally classified into three categories: location-based, 
range-based and connectivity-based. 

Location-based approaches are usually
based on computational geometry with
tools such as Voronoi diagram and Delaunay triangulations, to discover
coverage holes \cite{FGG04, WCL04, ZZF09}. Range-based approaches
attempt to discover coverage holes by using only relative
distances between neighbouring sensors \cite{B08, B12}.
These two types of approaches need either precise location information 
or accurate distance information, which restricts their applications since
such information is not easy to obtain in many settings.

In connectivity-based approaches, homology-based schemes attract particular
attention due to its powerfulness for coverage hole detection. As a pioneer work, in \cite{GM05}, Ghrist and his collaborators introduced homology to detect coverage holes. They first introduced a combinatorial object, $\check{\textrm{C}}$ech complex, which can capture all coverage holes. Unfortunately, this object is very difficult to construct even if the precise location information of sensors is provided. Thus, they introduced another more easily computable complex, Rips complex. 
This complex can be constructed with the sole knowledge of the connectivity graph of the network and gives an approximate coverage by simple algebraic calculations. Then their work is followed by \cite{DSG05, SG06, DSG07, DSG072}, where a relative homological criterion for coverage is presented. But these homology based approaches are  centralized. The first steps of implementing the above ideas in a distributed way were taken in \cite{ME06}. It is shown that combinatorial Laplacians are the right tools for distributed computation of homology groups, and thus can be used for decentralized coverage verification. The combinatorial Laplacians can be used to detect absence of holes or a single hole. But when there are multiple holes close to each other in WSNs, it is not clear how to distinguish them. To address such limitations, a gossip like decentralized algorithm was proposed in \cite{MJ07} to compute homology groups, but its convergence is slow as stated in \cite{MJ07}. In \cite{TJ10}, the authors first presented a decentralized scheme based on combinatorial Laplacians to verify whether there is a coverage hole or not in a WSN. For the case when there are coverage holes, they further formulated the problem of localizing coverage holes as an optimization problem for computing a sparse generator of the first homology group of the Rips complex corresponding to the sensor network. But it is possible that some cycle found by their algorithm contains multiple holes next to each other. More recently, for the purpose of coverage verification, a novel distributed algorithm for homology computation was proposed in \cite{DGJ11} based on reduction and co-reduction of simplicial complex. For the case when there are more than one holes, the authors proposed to first find the homology generator of the reduced complex and then use their algorithm to localize holes. The algorithm is quite original but with high complexity since it requires to construct all the simplices. All these homology based algorithms try to verify coverage or detect coverage holes by computing homology either in a centralized or distributed way, but computing homology is generally of high complexity especially for large scale sensor networks. Our algorithm does not try to compute homology to localize holes, it simplifies the Rips complex of a WSN by deleting some vertices and edges without changing homology and makes the Rips complex nearly planar. Then it can be efficient to detect coverage holes.

\section{Models and Definitions} \label{secmodel}

Consider a collection of stationary sensors (also called nodes)
deployed randomly in a planar target field. As usual, isotropic radio propagation is assumed. Each sensor monitors a region within a circle of radius $R_s$ and may communicate with other  sensors within a circle of radius $R_c$. Note that this assumption is mainly for analyzing the proportion of the area of triangular holes, it is not necessary for the algorithm proposed in Section \ref{secalg}.

In addition, some other assumptions are as follows.
\begin{enumerate}
\item There are sensors located on the external boundary of the target field. 
They are known as fence sensors and other sensors are referred to as internal sensors. 
Each fence sensor has two fence neighbours. This is also the general assumption
in many homology based algorithms \cite{DSG05, DSG07, GM05, TJ10}.
\item Although sensors are not aware of their locations, every sensor can know whether 
it is a fence or an internal node by using the mechanisms presented in \cite{B08} or 
other methods as in \cite{DLL10}. In fact, it is a conventional assumption adopted by
many existing range-based methods \cite{B08, KBS09} or connectivity methods \cite{TJ10, DLL10}.
\item Internal sensors are distributed in the planar target field according to 
a homogeneous PPP with intensity $\lambda$.
\item Each sensor has a unique ID.
\item The network has only one connected component.
\end{enumerate}

\begin{figure}[ht]
  \centering
  \includegraphics[width = 0.3\textwidth]{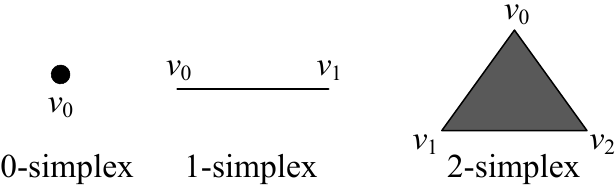}
  \caption{0-, 1- and 2-simplex}
  \label{simplex}
\end{figure}

Before defining $\check{\textrm{C}}$ech complex and Rips complex,
it is necessary to give a brief introduction to the tools used in the paper.
For further readings, see \cite{ARM83, mun84, HAT02}. 
Given a set of points $V$, a \textit{k}-simplex is an
unordered set $[v_0, v_1, ..., v_k] \subseteq V$ where $v_i \neq
v_j$ for all $i \neq j$. So a 0-simplex is a vertex, a 1-simplex is an
edge and a 2-simplex is a triangle with its interior included, see
Fig.  \ref{simplex}. The faces of this \textit{k}-simplex consist of
all (\textit{k}-1)-simplex of the form $[v_0, ..., v_{i-1},v_{i+1},
..., v_k]$ for $0 \leq i \leq k$. An abstract simplicial complex is a
collection of simplices which is closed with respect to inclusion of
faces.

Let $\mathcal{V}$ denote the set of
sensor locations in a WSN and $\mathcal{S}=\{s_v,\, v\in \mathcal{V}\}$  denote the collection of sensing
ranges of these sensors. For a location $v$,  $s_v = \{ x \in\mathbb{R}^2:\| x - v \| \leq R_s\}$. Then
$\check{\textrm{C}}$ech complex and Rips complex can be defined 
as follows \cite{DSG05,DSG07}.

\begin{definition}[$\check{\textrm{C}}$ech complex] 
  Given a finite collection of sensing ranges $\{s_v,\, v\in \mathcal{V}\}$,
  the $\check{\textrm{C}}$ech complex of the collection, $\check{\textrm{C}}(\mathcal{V})$, 
  is the abstract simplicial complex whose \textit{k}-simplices correspond to non-empty
  intersections of k + \emph{1} distinct elements of $\{s_v,\, v\in \mathcal{V}\}$.
\end{definition}

\begin{definition} [Rips complex]
  Given a finite set of points $\mathcal{V} $ in 
 $ \mathbb{R}^{n}$ and  a fixed radius
  $\epsilon$, the Rips complex of $\mathcal{V}$,
  $\mathcal{R}_\epsilon$($\mathcal{V}$), is the abstract simplicial complex
  whose k-simplices correspond to unordered \emph{(}k +
  \emph{1)}-tuples of points in $\mathcal{V}$ which are pairwise within
  Euclidean distance $\epsilon$ of each other.
\end{definition}

According to the definition, the
$\check{\textrm{C}}$ech complex and Rips complex of the WSN, respectively  denoted by $\check{\textrm{C}}_{R_s}
(\mathcal{V})$ and $\mathcal{R}_{R_c}(\mathcal{V})$, can be
 constructed as follows: a \textit{k}-simplex $[v_0, v_1,\cdots,v_k]$ belongs to $\check{\textrm{C}}_{R_s}
(\mathcal{V})$ whenever $\cap_{l=0}^k s_{v_l}\not =
\emptyset$ and a \textit{k}-simplex $[v_0, v_1,\cdots,v_k]$ belongs to
$\mathcal{R}_{R_c}(\mathcal{V})$ whenever $\|v_l-v_m\|\le R_c$ for all
$ 0\le l<m\le k$.

\begin{figure}[ht]
  \centering \subfloat[]{\includegraphics[width=0.14\textwidth]{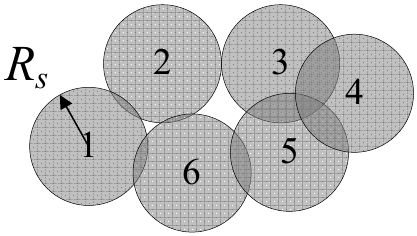}}
  \hfil
  \subfloat[]{\includegraphics[width=0.14\textwidth]{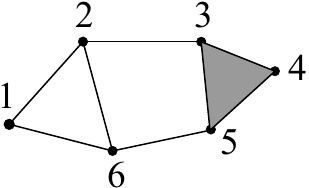}}
  \\
  \subfloat[]{\includegraphics[width=0.14\textwidth]{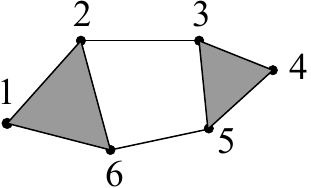}}
  \hfil
  \subfloat[]{\includegraphics[width=0.14\textwidth]{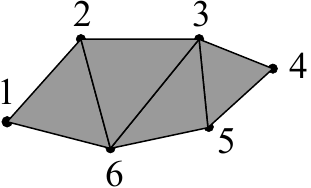}}
  \caption{(a) a WSN, (b) $\check{\textrm{C}}$ech complex, (c) Rips
    Complex under $R_c = 2 R_s$, (d) Rips Complex under $R_c = 2.5
    R_s$}
  \label{Rips}
\end{figure}

Fig. ~\ref{Rips} shows a WSN, its $\check{\textrm{C}}$ech complex and
two Rips complexes for two different values of $R_c$. Depending on the
ratio $R_c$ over $R_s$, the Rips complex and the $\check{\textrm{C}}$ech
complex may be close or rather different. In this example, for
$R_c=2R_s$, the Rips complex sees the hole surrounded by ${2,3,5,6}$
as in the $\check{\textrm{C}}$ech complex whereas it is missed in the
Rips complex for $R_c=2.5R_s$. At the same time, the true coverage
hole surrounded by ${1,2,6}$ is missed in both Rips complexes.

In fact, as proved in \cite{DSG05}, any coverage hole can be found in 
$\check{\textrm{C}}$ech complex. Unfortunately, the construction of $\check{\textrm{C}}$ech complex 
is of very high complexity even if the precise location information of nodes is provided. So a more easily computable tool, Rips complex, is used. But Rips complex can not always capture all coverage holes. To be more specific, there exist following relations between $\check{\textrm{C}}$ech complex and Rips complex:
\begin{equation}\label{cech rips}
 \mathcal{R}_{R_c}(\mathcal{V}) \subset \check{\textrm{C}}_{R_s}(\mathcal{V}) \subset \mathcal{R}_{2R_s}(\mathcal{V}), \hspace{4pt} \textrm{if} \hspace{2pt} R_c \leq \sqrt{3} R_s. 
\end{equation} 
According to (\ref{cech rips}), some relationships between $\check{\textrm{C}}$ech complex and Rips complex in terms of coverage hole can be derived as illustrated in the following corollaries. For convenience, define $\gamma = R_c / R_s $. 

\begin{corollary} \label{CRcase1}
When $\gamma \leq \sqrt{3}$, if there is no hole in Rips complex $\mathcal{R}_{R_c}(\mathcal{V})$, there must be no hole in $\check{\textrm{C}}$ech complex $\check{\textrm{C}}_{R_s}(\mathcal{V})$.
\end{corollary}

%\begin{IEEEproof}
%If there is no hole in $\mathcal{R}_{R_c}(\mathcal{V})$, it means that $\mathcal{R}_{R_c}(\mathcal{V})$ can be triangulated. Since $\gamma \leq \sqrt{3}$ means $R_c \leq \sqrt{3} R_s$, according to the first inclusion in (\ref{cech rips}), we can see that $\mathcal{R}_{R_c}(\mathcal{V}) \subset \check{\textrm{C}}_{R_s}(\mathcal{V})$. Consequently, $\check{\textrm{C}}$ech complex $\check{\textrm{C}}_{R_s}(\mathcal{V})$ can also be triangulated. And when $R_c \leq \sqrt{3} R_s$, each triangle must be covered by the sensing range of its vertex nodes \cite{DSG05}. So there is no hole in $\check{\textrm{C}}_{R_s}(\mathcal{V})$.
%\end{IEEEproof}

\begin{corollary} \label{CRcase2}
When $\gamma \geq 2$, if there is a hole in Rips complex $\mathcal{R}_{R_c}(\mathcal{V})$, there must be a hole in $\check{\textrm{C}}$ech complex $\check{\textrm{C}}_{R_s}(\mathcal{V})$.
\end{corollary}

%\begin{IEEEproof}
%If there is a hole in $\mathcal{R}_{R_c}(\mathcal{V})$, there must be a cycle with more than three edges in $\mathcal{R}_{R_c}(\mathcal{V})$ that can not be triangulated, as the cycle \{2, 3, 5, 6\} in Fig.  \ref{Rips}(c). Since $\gamma \geq 2$ means $R_c \geq 2 R_s$, according to the second inclusion in (\ref{cech rips}), we can see that $ \check{\textrm{C}}_{R_s}(\mathcal{V}) \subset \mathcal{R}_{2R_s}(\mathcal{V}) \subset \mathcal{R}_{R_c}(\mathcal{V})$. Consequently, there must also be a cycle in $\check{\textrm{C}}_{R_s}(\mathcal{V})$ which can not be triangulated. And there is a coverage hole in the cycle.
%\end{IEEEproof}

\begin{corollary} \label{CRcase3}
When $\sqrt{3} < \gamma < 2$, there is no guarantee relation between Rips complex $\mathcal{R}_{R_c}(\mathcal{V})$ and $\check{\textrm{C}}$ech complex $\check{\textrm{C}}_{R_s}(\mathcal{V})$ in terms of holes.
\end{corollary}

Please refer to \cite{YMD11} for the proof.

%\begin{IEEEproof}
%It is a direct corollary from Corollary \ref{CRcase1} and \ref{CRcase2}.
%\end{IEEEproof}

From the discussion above, a hole in a $\check{\textrm{C}}$ech complex 
not seen in a Rips complex must be bounded by a triangle. Based on this observation, 
a formal definition of 'triangular hole' and 'non-triangular hole' is given as follows.

\begin{definition}[Triangular and non-triangular hole]\label{trihole}
For a pair of complexes $\check{\textrm{C}}_{R_s}(\mathcal{V})$ and 
$\mathcal{R}_{R_c}(\mathcal{V})$, a triangular hole is an uncovered 
region bounded by a triangle formed by three nodes $v_0, v_1, v_2$, where 
$v_0, v_1, v_2$ can form a 2-simplex which appears in $\mathcal{R}_{R_c}(\mathcal{V})$ 
but not in $\check{\textrm{C}}_{R_s}(\mathcal{V})$. Other holes
are non-triangular.
\end{definition}

For triangular holes, it is impossible to detect them with only connectivity information,
so we want to analyze the proportion of the area of such holes in a target field. For 
non-triangular holes, we aim to design a distributed algorithm to discover the boundaries of 
these holes.

\section{Bounds on proportion of the area of triangular holes}

For triangular holes, we aim to derive the proportion of their area. In this section, the conditions under which any
point in the target field is inside a triangular hole are first given. In Section \ref{secmodel}, it is found that the proportion of the area of
triangular holes is related to the ratio $\gamma$. 
Three different cases are considered for the proportion computation. For each case, the upper and lower
bounds of the proportion are derived. 

\subsection{Preliminary} \label{secprelimi}

\begin{lemma} \label{condition} For any point in the target field, it
  is inside a triangular hole if and only if the following two conditions are 
  satisfied:
  \begin{enumerate}
  \item the distance between the point and its closest node is
    larger than $R_s$.
  \item the point is inside a triangle: the convex hull of three nodes with pairwise distance less than or equal to $R_c$.
  \end{enumerate}
\end{lemma}

\begin{figure}[ht]
  \centering
  \includegraphics[width=0.15\textwidth]{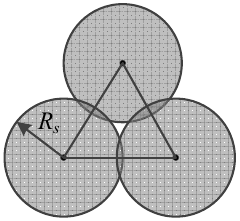}
  \caption{An example of a triangular hole.}
  \label{triholeexample}
\end{figure}

Fig.  \ref{triholeexample} gives an example to show a triangular hole. The blanket region inside the triangle is a triangular hole since it is not covered by any node and is bounded by a triangle.

\begin{lemma} \label{distance} If there exists a point $O$ which
  is inside a triangular hole, then $R_s<R_c/\sqrt{3}$. 
\end{lemma}

%\begin{IEEEproof}
%  According to the definition of triangular holes, if there exists
%  a triangular hole, then there must be a \textcolor{red}{2-simplex which is in $\mathcal{R}_{R_c}(\mathcal{V}$ but not in 
%  $\check{\textrm{C}}_{R_s}(\mathcal{V})$}.
%  If $R_s \geq R_c/\sqrt{3}$, then according to the first inclusion
%  in (1), we have $\mathcal{R}_{R_c}(\mathcal{V}) \subset \check{\textrm{C}}_{R_s}(\mathcal{V})$,
%  it means that there exists not a \textcolor{red}{2-simplex} which is in $\mathcal{R}_{R_c}(\mathcal{V})$
%  but not in $\check{\textrm{C}}_{R_s}(\mathcal{V})$, there is 
%  a contradiction, so $R_s<R_c/\sqrt{3}$.
%\end{IEEEproof}

\begin{lemma} \label{closedist} Let $O$ be a point inside a triangular hole
  and $l$ denote the distance between $O$ and its closest neighbour, then
  $R_s < l \leq R_c  / \sqrt{3}$.
\end{lemma}

Please refer to \cite{YMD12} for the proof.

%\begin{IEEEproof}
%  $R_s < l$ is a direct corollary from Lemma \ref{condition}. We only
%  need to prove $ l \leq R_c / \sqrt{3}$. If point $O$ is inside a
%  triangular hole, it must be surrounded by a triangle formed by
%  \textcolor{red}{three nodes} with pairwise distance less than or equal to $R_c$. Assume it is
%  surrounded by a triangle $\rm{N_0}\rm{N_1}\rm{N_2}$, as in Fig. 
%  \ref{triangle}. The closest neighbour of $O$ is not necessarily in
%  the set \{$\rm{N_0}, \rm{N_1}, \rm{N_2}$\}. If $l > R_c / \sqrt{3}$,
%  then $d_0 \geq l > R_c / \sqrt{3}$, $d_1 \geq l > R_c / \sqrt{3}$
%  and $d_2 \geq l > R_c / \sqrt{3}$. In addition, since \textcolor{red}{$\alpha + \beta + \varphi = 2\pi$}, there must
%  be one angle no smaller than $2\pi/3 $. \textcolor{red}{Without loss of generality,} assume \textcolor{red}{$\alpha \geq 2\pi/3$}. Then according to the law
%  of cosines, $d_{02}^2 = d_0^2 + d_2^2 - 2 d_0 d_2 \cos\alpha >
%  R_c^2/3 + R_c^2/3 - 2/3 R_c R_c\cos(2\pi/3) = R_c^2$. So
%  $d_{02}>R_c$. Since $\rm{N_0}$ and $\rm{N_2}$ are neighbours,
%  $d_{02}\leq R_c$. There is a contradiction. Therefore $l \leq R_c /
%  \sqrt{3}$.
%\end{IEEEproof}

%\begin{figure}[ht]
%  \centering
%  \includegraphics[width=0.3\textwidth]{triangle}
%  \caption{Illustration of $O$ being inside a triangular hole}
%  \label{triangle}
%\end{figure}

A PPP whose intensity is proportional to the
Lebesgue measure is stationary in the sense that any translation of
its atoms by a fixed vector does not change its law. Thus without 
considering border effect \cite{BZ02}, any point
has the same probability to be inside a triangular hole as the
origin $O$. This probability in a homogeneous setting is also equal to
the proportion of the area of triangular holes. We borrow part of the line of proof from \cite{LHZ12}
where a similar problem is analyzed.

We consider the probability that the origin $O$ is inside a
triangular hole. Since the length of each edge in the Rips complex must be
at most $R_c$, only the nodes within $R_c$ from the origin can
contribute to the triangle which bounds a triangular hole containing the
origin. Therefore, we only need to consider the PPP
constrained in the closed ball $B(O, R_c)$ which is also a
homogeneous PPP with intensity $\lambda$. We denote this
process as $\Phi$. In addition, $T(x, y, z)$ denotes the property that
the origin $O$ is inside the triangular hole
bounded by the triangle with points $x, y, z$ as vertices. When $n_0, n_1,
n_2$ are points of the process $\Phi$, $T(n_0, n_1, n_2)$ is also used
to denote the event that the triangle formed by the nodes $n_0,
n_1, n_2$ bounds a triangular hole containing the origin. In addition,
we use $T'(n_0, n_1, n_2)$ to denote the event that the nodes
$n_0, n_1, n_2$ can not form a triangle which bounds a triangular hole 
containing the origin. 

\begin{figure}[ht]
  \centering
  \includegraphics[width=0.25\textwidth]{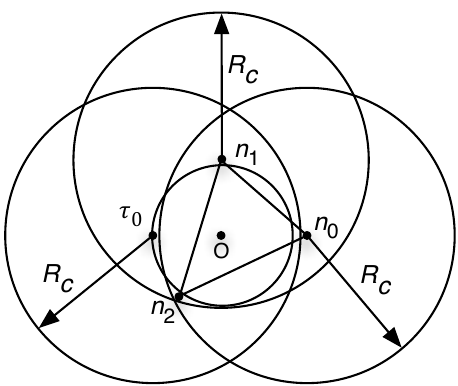}
  \caption{An example showing that the node closest to the origin $\tau_0$ does not contribute to a triangle which bounds a triangular hole containing the origin because the distance to $n_0$ is larger than $R_c$ and the triangle formed by $\tau_0, n_1, n_2$ does not contain $O$. In contrast, $n_0, n_1, n_2$ can form a triangle which bounds a triangular hole containing $O$. Here we assume the distance between $\tau_0$ and $O$ is larger than $R_s$. }
  \label{triholecase2}
\end{figure}

Let $\tau_0 = \tau_0(\Phi)$ be the node in the process $\Phi$ which is
closest to the origin. There are two cases for the origin to be inside a
triangular hole. The first case is that the node $\tau_0$ can contribute to
a triangle which bounds a triangular hole containing the origin. The
second case is that the node $\tau_0$ can not contribute to any triangle
which bounds a triangular hole containing the origin but other three
nodes can form a triangle which bounds a triangular hole containing the 
origin, as illustrated in Fig.  \ref{triholecase2}. So the probability that the origin is inside a triangular hole 
can be defined as 
\begin{equation} \label{eq0}
  \begin{split}
    &p(\lambda, \gamma) = \mathrm{P}\{O \textrm{ is inside a triangular hole}\} \\
    &= \mathrm{P}\{\bigcup_{\{n_0, n_1, n_2\} \subseteq \Phi} T(n_0, n_1, n_2)\} \\
    &= \mathrm{P}\{\bigcup_{\{n_1, n_2\} \subseteq \Phi
      \backslash \{\tau_0(\Phi)\}} T(\tau_0, n_1, n_2)\} + p_{sec}(\lambda, \gamma)
  \end{split}
\end{equation}
where 
\begin{equation*}
p_{sec}(\lambda, \gamma) = \mathrm{P}\{\bigcup_{\{n_0,\cdots, n_4\} \atop \subseteq \Phi
            \backslash \{\tau_0(\Phi)\}} T(n_0, n_1, n_2)\mid T^\prime(\tau_0, n_3, n_4) \}
\end{equation*} denotes the probability of the second case. $p_{sec}(\lambda, \gamma)$
is generally very small and obtained by simulations.

In the following parts, we will analyze this probability in three different cases.

\subsection{Case $0 < \gamma \leq \sqrt{3}$}

\begin{theorem} \label{trihole1} When $0 < \gamma \leq \sqrt{3}$, 
	$p(\lambda, \gamma) = 0$.
\end{theorem}

\begin{IEEEproof}
  It is a direct corollary from Lemma \ref{distance}.
\end{IEEEproof}

\subsection{Case $\sqrt{3} < \gamma \leq 2 $ } \label{seccase2}

\begin{theorem} \label{trihole2}
	When $\sqrt{3} < \gamma \leq 2 $, $p_l(\lambda, \gamma) < p(\lambda, \gamma) < p_u(\lambda, \gamma)$,
	where 
	\begin{equation} \label{lower1}
	  \begin{split}
	  	    p_l (&\lambda , \gamma) =  2\pi\lambda ^2 \int_{R_s}^{R_c/\sqrt{3}} r_0 dr_0 \int_{\alpha_0}^{\alpha_1} d\theta_1 \int_{r_0}^{R_1(r_0, \theta_1)} \\
	  	    & e^{-\lambda \pi r_0^2} \times e^{-\lambda |S^+(r_0, \theta_1)|}
	  	    (1 - e^{-\lambda |S^-(r_0, r_1, \theta_1)|})r_1 dr_1
	  \end{split}
	\end{equation}
	and
	\begin{equation} \label{upper1}
	  \begin{split}
	  	    p_u (&\lambda , \gamma)  = 2\pi\lambda ^2 \int_{R_s}^{R_c/\sqrt{3}} r_0 dr_0 \int_{\alpha_0}^{\alpha_1} d\theta_1 \int_{r_0}^{R_1(r_0, \theta_1)} \\
	  	    & e^{-\lambda \pi r_0^2} \times e^{-\lambda |S^+(r_0, \theta_1)|} (1 - e^{-\lambda |S^-(r_0, r_0, \theta_1)|})r_1 dr_1\\
	  	    & + p_{sec}(\lambda, \gamma)
	  \end{split}
	\end{equation}
	and
	\begin{align*}
		  & \alpha_0 = 2\arccos(R_c /(2r_0))\\
		  & \alpha_1 = 2\arcsin(R_c /(2r_0)) - 2\arccos(R_c/(2r_0))\\
		  & R_1(r_0, \theta_1) = \min(\sqrt{R_c^2 - r_0^2\sin^2 \theta_1} - r_0 \cos \theta_1,\\
		  & \qquad \qquad \sqrt{R_c^2 - r_0^2\sin^2 (\theta_1+\alpha_0)} + r_0 \cos (\theta_1+\alpha_0)) \\
		  & |S^+(r_0, \theta_1)| = \int_{\alpha_0} ^{\theta_1} \int_{r_0}^{R_1(r_0, \theta)} r dr d\theta \\
		  & |S^-(r_0, r_1, \theta_1)| = \int_{\theta_{2l}}^{-\alpha_0} \int_{r_0}^{R_2(r_0, r_1, \theta_1, \theta_2)}r dr d\theta_2\\
		  & \theta_{2l} = \theta_1 - \arccos \frac{\cos (R_c/R) - \cos \theta_1 \cos \theta_0}{\sin \theta_1 \sin \theta_0} \\
		  	& R_2(r_0, r_1, \theta_1, \theta_2) = \min(\sqrt{R_c^2 - r_0^2\sin^2 \theta_2} - r_0 \cos \theta_2, \\
		  	& \qquad \qquad \sqrt{R_c^2 - r_1^2\sin^2 (\theta_2-\theta_1)} + r_1 \cos (\theta_2-\theta_1) )
	\end{align*}
$p_{sec}(\lambda, \gamma)$ is obtained by simulations.
\end{theorem}

Since the proof is tedious, we only give the idea and main steps here.
See Appendix \ref{app1} for detailed computation.

For the lower bound, we only consider 
the first case that the closest node $\tau_0$ can contribute to a
triangle which bounds a triangular hole containing the origin.
The main idea is to first fix the closest node $\tau_0$, and then sequentially
decide the regions where the other two nodes may lie in, and finally
do a triple integral.

\begin{figure}[ht]
  \centering
  \includegraphics[width=0.35\textwidth]{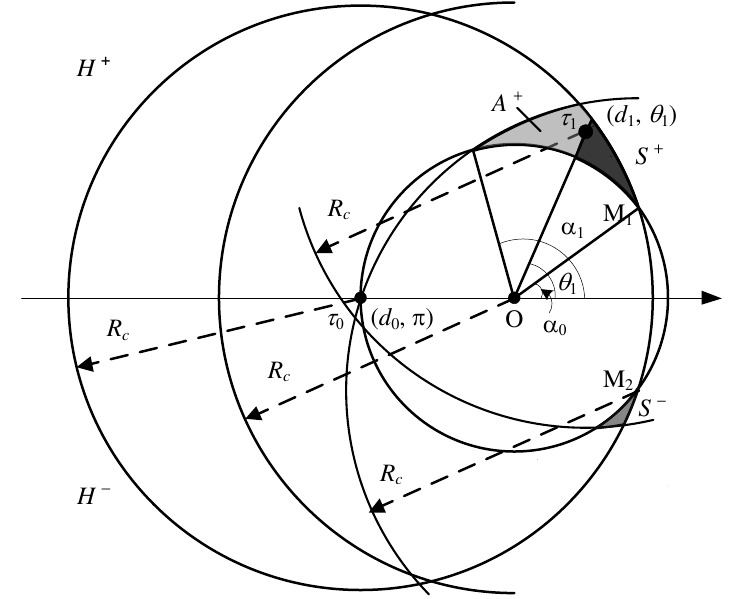}
  \caption{Illustration of regions $A^+, S^+$ and $S^-$ in the case $\sqrt{3} < \gamma \leq 2 $ }
  \label{case2lower2}
\end{figure}

Using polar coordinates, we assume the closest node $\tau_0$ lies on $(d_0, \pi)$. Once the node $\tau_0$ is determined, the other two nodes must lie in 
the different half spaces: one in $H^+ = \mathbb{R}^+ \times (0, \pi) $ 
and the other in $H^- = \mathbb{R}^+ \times (-\pi, 0) $. 
Assume $n_1$ lies in $H^+$ and $n_2$ lies in $H^-$.
Since the distance to $\tau_0$ is at most $R_c$,
$n_1$ and $n_2$ must also lie in the ball $B(\tau_0, R_c)$. Furthermore, the
distance to the origin is at most $R_c$ and larger than $d_0$, they
should also lie in the region $B(O, R_c) \backslash B(O, d_0)$. Therefore, 
$n_1$ must lie in $H^+ \bigcap B(\tau_0, R_c) \bigcap B(O, R_c)
\backslash B(O, d_0) $ and $n_2$ must lie in $H^- \bigcap
B(\tau_0, R_c) \bigcap B(O, R_c) \backslash B(O, d_0)$. In addition,
considering the distance between $n_1$ and $n_2$ should be at most
$R_c$ and the origin should be inside the triangle formed by $\tau_0$,
$n_1$ and $n_2$, $n_1$ must lie
in the shadow region $A^+ = H^+ \bigcap B(\tau_0, R_c) \bigcap B(O, R_c)
\backslash B(O, d_0) \bigcap B(M_2, R_c)$, shown in Fig.  \ref{case2lower2}. 
$M_2$ is one intersection point between the circle $C(O, d_0)$ and
the circle $C(\tau_0, R_c)$, such intersection point must exist 
in this case since $R_c = \gamma R_s \leq 2R_s < 2d_0$.

Ordering the nodes in $A^+$ by increasing polar angle so that $\tau_1 = (d_1, \theta_1)$ 
has the smallest angle $\theta_1$. And assume the nodes
$\tau_0$, $\tau_1$ and another node $\tau_2 \in H^- \bigcap B(\tau_0, R_c) \bigcap B(0, R_c)
\backslash B(0, d_0)$ can form a triangle which bounds a triangular hole containing
the origin, then $\tau_2$ must lie to the right of the line passing
through $\tau_1$ and $O$, denoted by $H^+(\theta_1)$ which contains
all points with polar angle $\theta \in (\theta_1 - \pi, \theta_1)$. In addition, the
distance to $\tau_1$ is less than $R_c$. So the node $\tau_2$ must lie
in the region $S^-$, as illustrated in Fig.  \ref{case2lower2}.
\begin{equation*}
  \begin{split}
    & S^-(\tau_0, \tau_1)  =  S^- (d_0, d_1, \theta_1) =  H^- \bigcap B(\tau_0, R_c)  \\
    & \bigcap B(0, R_c) \backslash B(0, d_0) \bigcap H^+(\theta_1)
    \bigcap B(\tau_1, R_c)
  \end{split}
\end{equation*}

\begin{figure}[ht]
  \centering
  \includegraphics[width=0.35\textwidth]{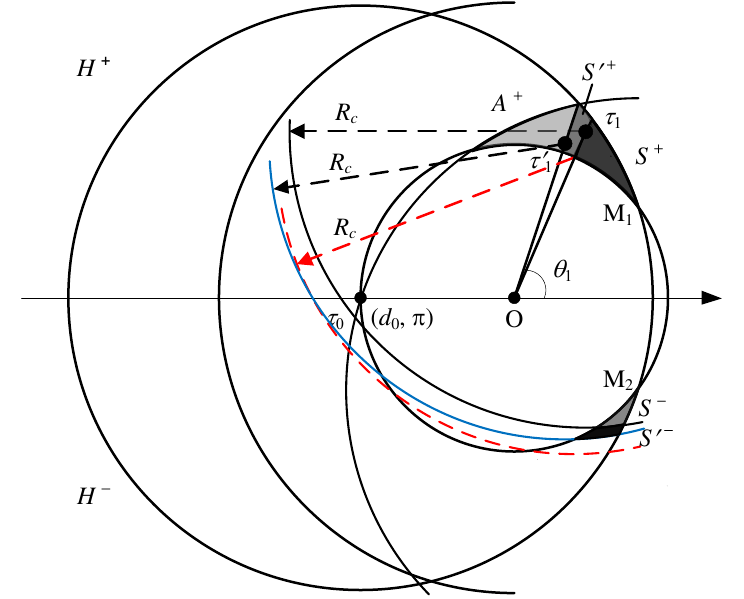}
  \caption{Illustration of regions $S^{\prime+}$ and $S^{\prime-}$ in the case $\sqrt{3} < \gamma \leq 2 $ }
  \label{case2upper}
\end{figure}

Assume only $\tau_0, \tau_1$ and nodes in $S^-(\tau_0, \tau_1)$ can
contribute to the triangle which bounds a triangular hole containing the
origin, we can get a lower bound of the probability that the
origin is inside a triangular hole. It is a lower bound because it is
possible that $\tau_1$ can not contribute to a triangle which bounds a 
triangular hole containing the origin, but some other nodes with 
higher polar angles in the region $A^+$ can contribute to such a triangle.
E.g. in Fig.  \ref{case2upper}, if there is no node in $S^-$ but 
there are some nodes in $S^{\prime-}$, then $\tau_1$ can not contribute
to any triangle which bounds a triangular hole containing the origin, but
$\tau_1^\prime$ can form such a triangle with $\tau_0$ and another node in
$S^{\prime-}$. 

%\begin{equation}
%  \begin{split}
%    p(\lambda &, \gamma) > p_l (\lambda, \gamma) \\
%    =& 2\pi\lambda ^2 \int_{R_s}^{R_c/\sqrt{3}} r_0 dr_0 \int_{2\arccos \frac{R_c}{2r_0}}^\pi d\theta_1 \int_{r_0}^{R_1(r_0, \theta_1)} \\
%    & e^{-\lambda \pi r_0^2} \times e^{-\lambda |S^+(r_0, \theta_1)|}
%    (1 - e^{-\lambda |S^-(r_0, r_1, \theta_1)|})r_1 dr_1
%  \end{split}
%\end{equation}

Next we prove the upper bound. As discussed in Section \ref{secprelimi},
there are two cases for the origin being inside a triangular hole. As for the 
second case that the closest node $\tau_0$ can not but some other nodes
can contribute to a triangle which bounds a triangular hole containing
the origin, it is not easy to obtain a closed-form expression for such
probability, we can get it by simulations. Simulation results show
that this probability is less than 0.16\% at any $\gamma \leq 3$ with any 
intensity $\lambda$. So we still focus on the probability of the first
case.

Still consider the nodes in $A^+$, each node $(d, \theta)$
corresponds to an area $|S^-(d_0, d, \theta)|$. The higher is the area $|S^-(d_0, d, \theta)|$, the higher is the probability that there is at least one node in
$S^-(d_0, d, \theta)$, consequently the probability of the first case
will be higher. It can be seen from Fig.  \ref{case2lower2} that the closer
to $\theta_1$ is $\theta$ and the closer to $d_0$ is $d_1$, the higher is 
the area $|S^-(d_0, d, \theta)|$, so the largest area $|S^-(d_0, d, \theta)|$
is $|S^-(d_0, d_0, \theta_1)|$. Based on that, the upper bound can be derived.

As can be seen, the expression for lower bound is closed-form, while the expression for upper bound is not 
exactly closed-form since it includes a non-analytical part $p_{sec}(\lambda, \gamma) $. For lower bound and the 
closed-form part for upper bound, we use numerical integration to approximate the triple integrals. For 
$p_{sec}(\lambda, \gamma) $, we get it by simulations since it is generally very small, it has little impact on the derived upper bound.

%\begin{equation}
%  \begin{split}
%    & p(\lambda, \gamma) < p_u (\lambda, \gamma) \\
%    & = 2\pi\lambda ^2 \int_{R_s}^{R_c/\sqrt{3}} r_0 dr_0 \int_{2\arccos \frac{R_c}{2r_0}}^\pi d\theta_1 \int_{r_0}^{R_1(r_0, \theta_1)} \\
%    & e^{-\lambda \pi r_0^2} \times e^{-\lambda |S^+(r_0, \theta_1)|}
%    (1 - e^{-\lambda |S^-(r_0, r_0, \theta_1)|})r_1 dr_1
%  \end{split}
%\end{equation}

\subsection{Case $\gamma > 2 $}

\begin{theorem} \label{trihole3}
	When $\gamma > 2 $, $p_l(\lambda, \gamma) < p(\lambda, \gamma) < p_u(\lambda, \gamma)$,
	where 
	\begin{equation} \label{lower2}
	  \begin{split}
	    p_l &(\lambda, \gamma) =  2\pi\lambda ^2 \Big \{ \int_{R_s}^{R_c/2} r_0 dr_0 \int_{0}^\pi d\theta_1 \int_{r_0}^{R_1^{'}(r_0, \theta_1)} \\
	    &  e^{-\lambda \pi r_0^2} \times e^{-\lambda |S^+(r_0, \theta_1)|} (1 - e^{-\lambda |S^-(r_0, r_1, \theta_1)|})r_1 dr_1 \\
	    & + \int_{R_c/2}^{R_c/\sqrt{3}} r_0 dr_0 \int_{\alpha_0}^{\alpha_1} d\theta_1 \int_{r_0}^{R_1(r_0, \theta_1)} e^{-\lambda \pi r_0^2}\\
	    & \times e^{-\lambda |S^+(r_0, \theta_1)|} (1 - e^{-\lambda
	      |S^-(r_0, r_1, \theta_1)|})r_1 dr_1 \Big\}
	  \end{split}
	\end{equation}
	and 
	\begin{equation} \label{upper2}
	  \begin{split}
	    p_u &(\lambda, \gamma) =  2\pi\lambda ^2 \Big \{ \int_{R_s}^{R_c/2} r_0 dr_0 \int_{0}^\pi d\theta_1 \int_{r_0}^{R_1^{'}(r_0, \theta_1)} \\
	    &  e^{-\lambda \pi r_0^2} \times e^{-\lambda |S^+(r_0, \theta_1)|} (1 - e^{-\lambda |S^-(r_0, r_0, \theta_1)|})r_1 dr_1 \\
	    & + \int_{R_c/2}^{R_c/\sqrt{3}} r_0 dr_0 \int_{\alpha_0}^{\alpha_1} d\theta_1 \int_{r_0}^{R_1(r_0, \theta_1)} e^{-\lambda \pi r_0^2}\\
	    & \times e^{-\lambda |S^+(r_0, \theta_1)|} (1 - e^{-\lambda
	      |S^-(r_0, r_0, \theta_1)|})r_1 dr_1 \Big\}\\
	    & + p_{sec}(\lambda, \gamma)
	  \end{split}
	\end{equation}
	and
	\begin{equation*}
	 \begin{split}
	    R_1^{'}(r_0, \theta_1)& = \min(\sqrt{R_c^2 - r_0^2\sin^2 \theta_1} - r_0 \cos \theta_1,\\
	   		  	    & \qquad \qquad \sqrt{R_c^2 - r_0^2\sin^2 \theta_1} + r_0 \cos \theta_1)
	 \end{split}
	\end{equation*}
	$p_{sec}(\lambda, \gamma)$ is obtained by simulations.
\end{theorem}

\begin{figure}[ht]
  \centering \subfloat[]{\includegraphics[width=0.35\textwidth]{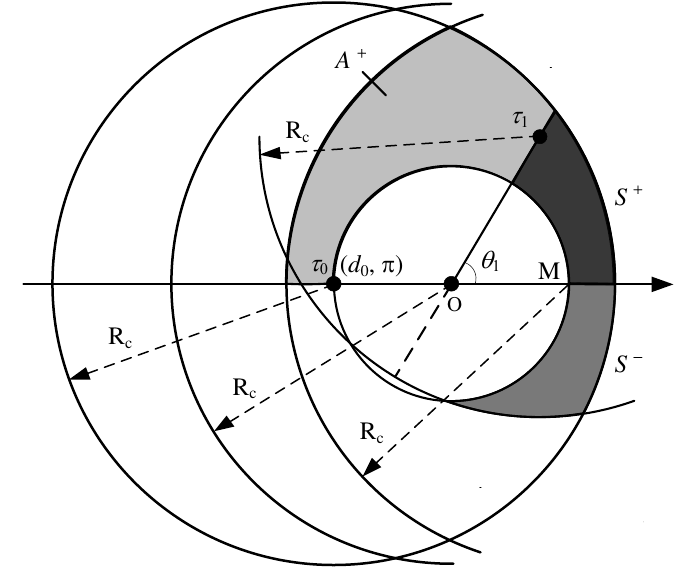}}
  \\
  \subfloat[]{\includegraphics[width=0.35\textwidth]{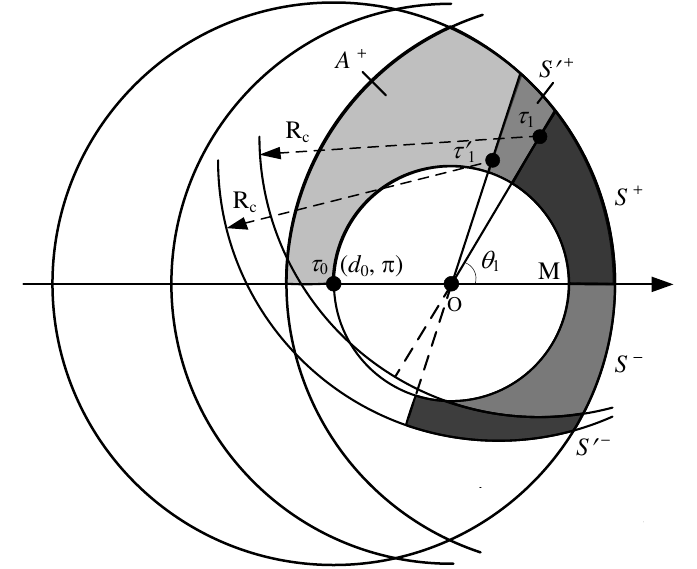}}
  \caption{Illustrations of regions in case $\gamma > 2$. (a) the regions $A^+,
   S^+$ and $S^-$ (b) the regions $S^{\prime+}$ and $S^{\prime-}$ }
  \label{case3area}
\end{figure}

In this case, we can use the same method as in Section \ref{seccase2} to
get the lower and upper bounds, shown in (\ref{lower2}) and (\ref{upper2})
respectively. But we need to consider two situations
$R_s < d_0 \leq R_c/2$ and $R_c/2 < d_0 \leq R_c/\sqrt{3}$. In the
first situation, $d_0 \leq R_c/2$ means that the ball $B(O, d_0)$ is 
included in the ball $B(\tau_0, R_c)$. The illustrations for the regions 
$A^+, S^+, S^-, S^{\prime+}$ and $S^{\prime-}$ are shown in 
Fig.  \ref{case3area}.
In addition, the lower limit of integration for $\theta_1$ is 0 and
the upper limit is $\pi$. The second
situation is the same as that in Section \ref{seccase2}.

\section{Distributed coverage hole detection algorithm} \label{secalg}

For non-triangular holes, we aim to design an efficient distributed algorithm to 
detect their minimum boundary cycles. The basic idea is that for the Rips
complex of a WSN, we try to delete some vertices and edges without
changing the homology while making the Rips complex more sparse and
nearly planar. Then it is easy to find boundary edges (1-simplices), each of which has at most one neighbour. Finally such edges are connected in some order to form boundary cycles.

\begin{figure}[ht]
 \centering
 \includegraphics[width = 0.25\textwidth]{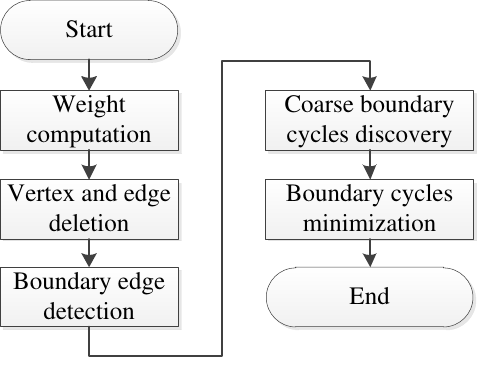}
 \caption{Flow chart of the algorithm}
 \label{flow}
\end{figure}

\begin{figure*}[!t]
\centering
\subfloat[]{\includegraphics[width=0.24\textwidth]{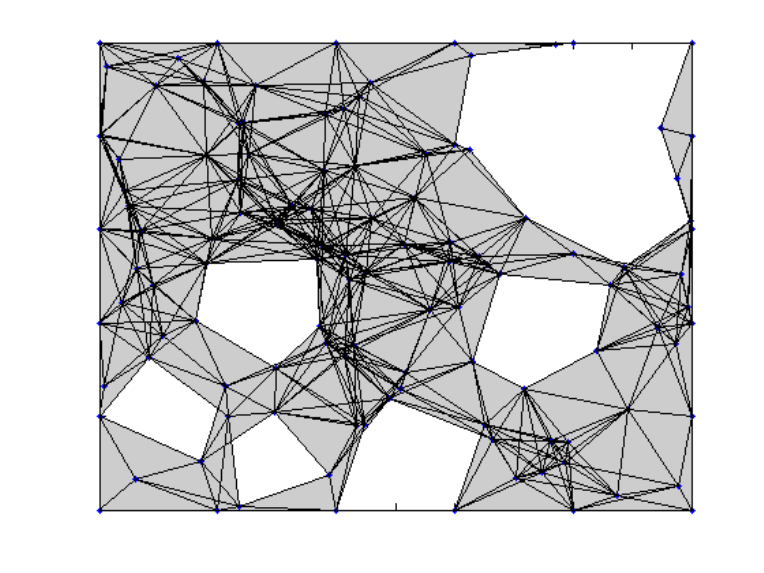}}
\hfil
\subfloat[]{\includegraphics[width=0.24\textwidth]{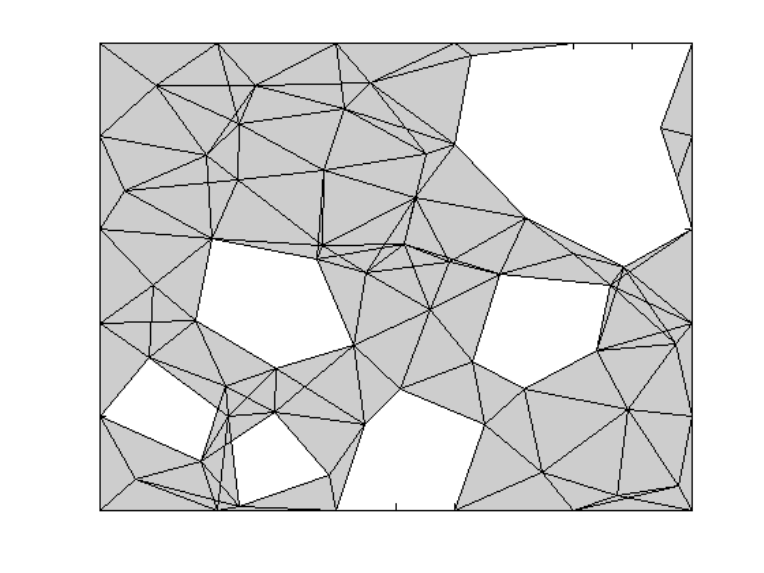}}
\hfil
\subfloat[]{\includegraphics[width=0.24\textwidth]{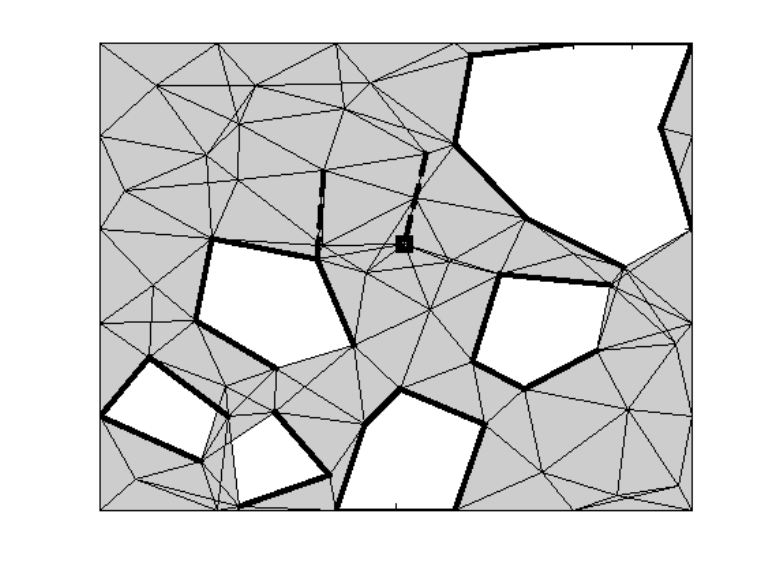}}
\hfil
\subfloat[]{\includegraphics[width=0.24\textwidth]{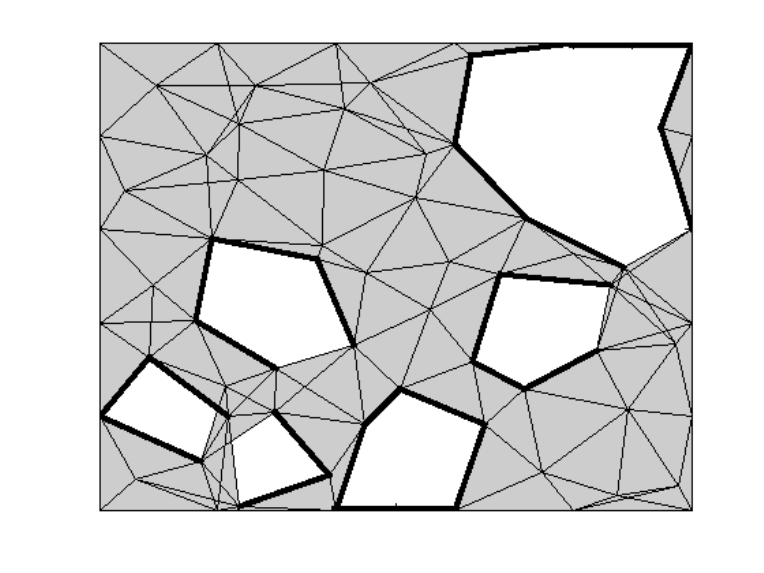}}
\\
\subfloat[]{\includegraphics[width=0.24\textwidth]{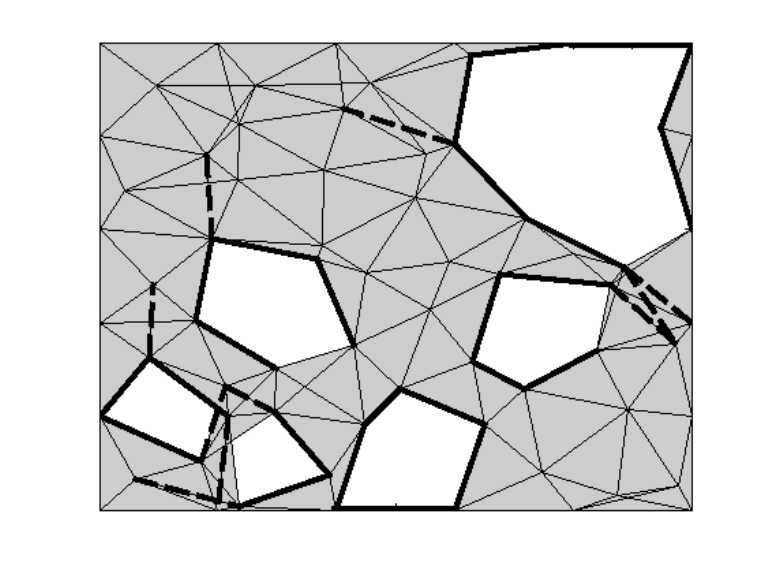}}
\hfil
\subfloat[]{\includegraphics[width=0.24\textwidth]{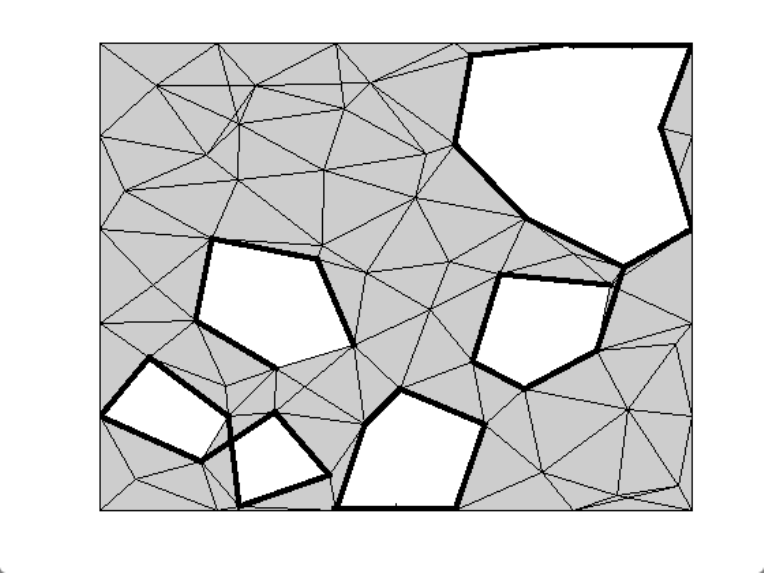}}
\hfil
\subfloat[]{\includegraphics[width=0.24\textwidth]{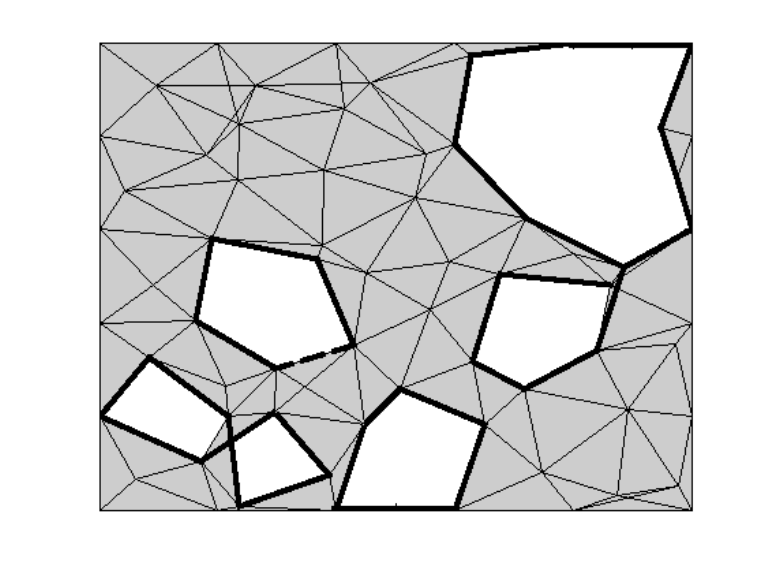}}
\hfil
\subfloat[]{\includegraphics[width=0.24\textwidth]{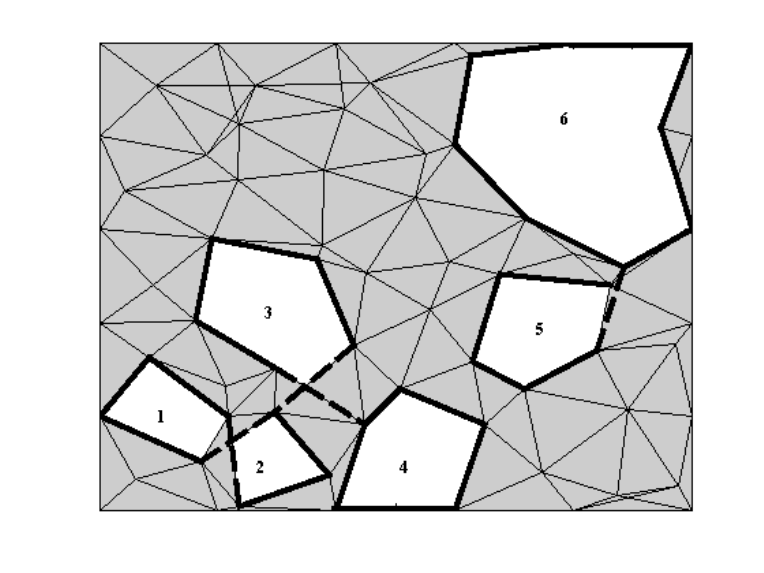}}
\\
\subfloat[]{\includegraphics[width=0.24\textwidth]{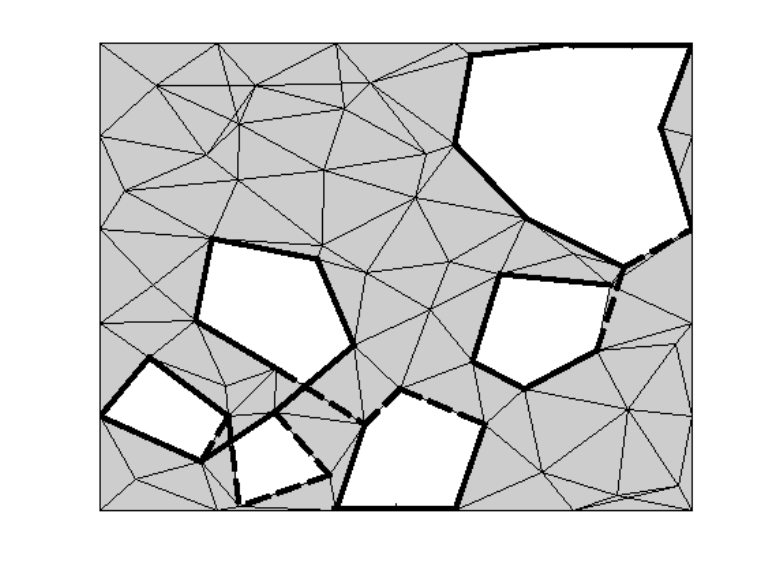}}
\hfil
\subfloat[]{\includegraphics[width=0.24\textwidth]{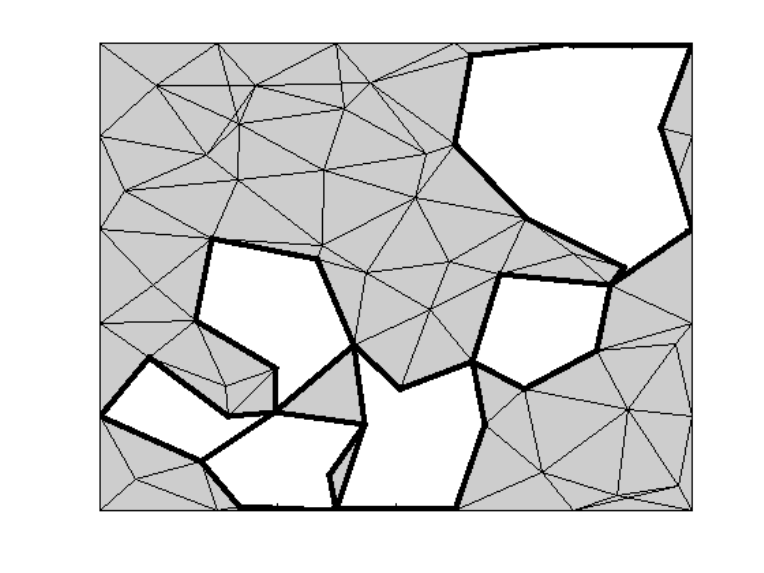}}
\hfil
\subfloat[]{\includegraphics[width=0.24\textwidth]{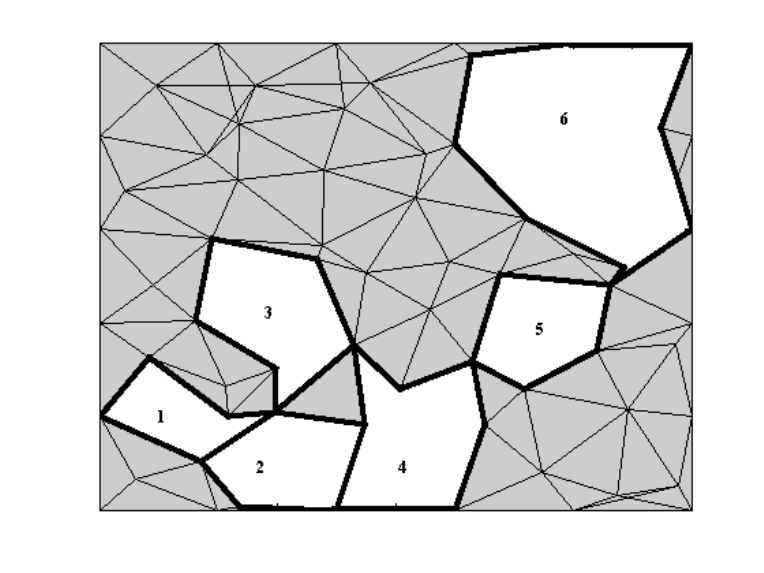}}
\hfil
\subfloat[]{\includegraphics[width=0.24\textwidth]{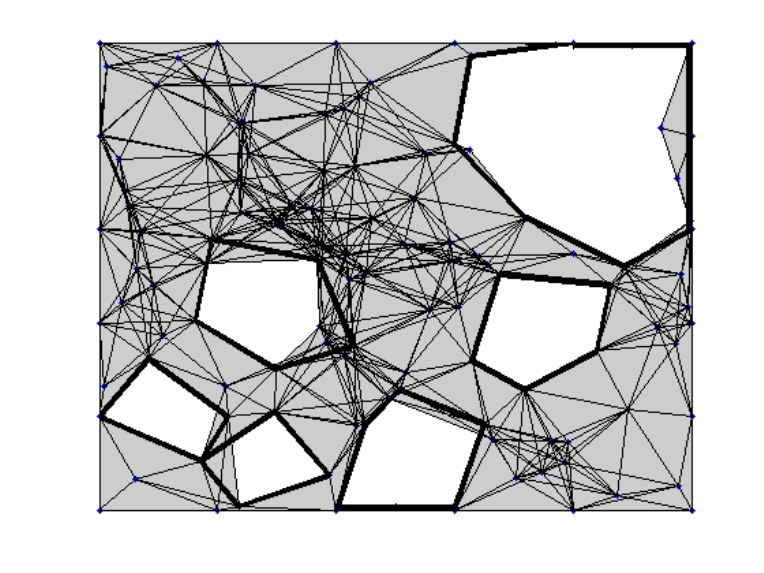}}
\caption{Procedures of the boundary detection algorithm. (a) Rips complex of a WSN, 
(b) after vertex deletion, (c$\sim$d) edge deletion, (e$\sim$j) boundary edge detection,
(k) coarse boundary cycles discovery, (l) boundary cycles minimization}
\label{procedure}
\end{figure*}

More specifically, our algorithm includes five components: weight computation, vertex
and edge deletion, boundary edge detection, coarse boundary cycles 
discovery and boundary cycles minimization, as shown in Fig.  \ref{flow}. 
An example is used to illustrate the procedures of this algorithm 
in Fig.  \ref{procedure}. 
In weight computation component, the Rips complex
of the WSN is first constructed, shown in Fig.  \ref{procedure}(a), then 
each node computes its weight independently.
The definition of weight of a node will be presented in the next part.
After obtaining the weight, each node continues to determine whether it
is deletable or not according to some rule defined hereafter. 
Fig.  \ref{procedure}(b) shows the result of vertex deletion.
Furthermore, some special edges may be deleted. Fig.  \ref{procedure}(c)
shows the process of such special edge deletion. After the second
component, many boundary edges can be found, as the bold lines shown in Fig.  \ref{procedure}(d). 
But it is possible that some 
other boundary edges have not been found. Then in the third component, all 
or nearly all boundary edges will be found after deleting some edges, see Fig.  \ref{procedure}(e$\sim$j). 
Then coarse boundary cycles can be easily discovered, as shown in Fig.  \ref{procedure}(k). 
The found boundary cycles may not be minimum. 
In this case, coarse boundary cycles will be minimized in the final 
component as shown in Fig.  \ref{procedure}(l).

\subsection{Definitions} \label{secdefinition}

%Before presenting the details of our algorithm, we first give some definitions as follows.

We say that a $i$-simplex $[v_{i0}, v_{i1}, ..., v_{ii}]$is part of a $j$-simplex
$[v_{j0}, v_{j1}, ..., v_{jj}]$ if $[v_{i0}, v_{i1}, ..., v_{ii}] \subset [v_{j0}, v_{j1}, ..., v_{jj}]$.
So the vertex $[v_0]$ and $[v_1]$ is part of the edge $[v_0, v_1]$. The edge
$[v_0, v_1]$ is part of the triangle $[v_0, v_1, v_2]$. In addition, we use
$E(v)$ to denote all the edges that the node $v$ is part of and $T(v)$
to denote all the 2-simplices that the node $v$ is part of.

\begin{definition}[Index of a 2-simplex]
The index of a \emph{2}-simplex $\triangle$ is the highest dimension of the simplex that the \emph{2}-simplex is part of, denoted by $I_\triangle$.
\end{definition}

\begin{definition}[Weight of a node]
The weight of a fence node is defined to be \emph{0}.
For any internal node $v$, if there exists one edge in $E(v)$ which is not
part of any \emph{2}-simplex, the weight $w_v$ of node $v$ is set to be \emph{0}; if not,
the weight is the minimum index of all the \emph{2}-simplices in $T(v)$, that is
$w_v = \min_{\triangle \in T(v)} {I_\triangle}$
\end{definition}

The weight of an internal node is an indicator of the density of surrounding nodes.
If the weight of an internal node is 0, the node must be on the boundary of 
a coverage hole. The larger the weight is, the higher is the probability that the node
is not on the boundary of a coverage hole. 

We also use the definition of simple-connectedness graph as in \cite{DLL09}.
Let $G$ be a simple graph with vertex set $V(G)$ and edge set $E(G)$. A cycle
$C$ is a sub-graph of $G$ if it is connected and each vertex in $C$ has degree two.
The length of a cycle $C$ is the number of its edges, denoted by $|E(C)|$.
The cycle space $C(G)$ of a graph $G$ contains all the cycles in $G$. The addition 
of two cycles $C_1$ and $C_2$ is defined as $C_1 \oplus C_2 = (E(C_1) \cup E(C_2)) \setminus (E(C_1) \cap E(C_2))$.
The triangle cycle subspace $C_T(G)$ of G is the set of all 3-length cycles in $C(G)$.

\begin{definition}[Simple-Connectedness Graph]
A connected graph $G$ is of simple connectedness if its cycle space $C(G)$ is empty, or
for any cycle $C$ in $C(G)$, there exists a set of \emph{3}-length cycles $T_0 \subseteq C_T(G)$ 
such that $C = \sum _{T \in T_0} T$, which means all cycles in $C(G)$ can be triangulated.
\end{definition}

Let $X$ be a vertex (or edge) set in a graph $G$, we use $G[X]$ to denote the vertex-induced
(or edge-induced) sub-graph by $X$. The neighbour set of a vertex $v$ in $G$ is denoted by $N_G(v)$.
The neighbouring graph $\Gamma_G(v)$ of vertex $v$ is denoted as $G[N_G(v)]$. The neighbouring
graph  $\Gamma_G(e)$ of an edge $e = [u, v]$ is defined as $G[N_G(u) \cap N_G(v) \cup\{u, v\}] - e$. 
The neighbour 
set of $k$-simplex $[v_0, v_1, ..., v_k]$ is defined as $\bigcap_{i=0}^k N_G(v_i)$.

\begin{definition} [Deletion of a $k$-simplex in Rips complex $\mathcal{R}(\mathcal{V})$]
  A $k$-simplex $[v_0, v_1, \cdots, v_k]$ is deleted in a Rips complex $\mathcal{R}(\mathcal{V})$
  means that the simplex and all simplices which the simplex is part of are deleted from $\mathcal{R}(\mathcal{V})$.
\end{definition}

Based on definitions above, we can give the definition of HP (Homology Preserving) transformation. 

\begin{definition}[HP Transformation]
A HP transformation is a sequential combination of vertex (or edge) deletion as follows:
a vertex (or edge) $x$ of $G$ is deletable if neighbouring graph $\Gamma_G(x)$ 
\emph{(1)} has two or more vertices; \emph{(2)} is connected and \emph{(3)} is a simple-connectedness graph.
\end{definition}

\begin{theorem}
HP transformations do not change the number of coverage holes in Rips complex of a WSN.
\end{theorem}

\begin{IEEEproof}
  In order to prove HP transformations do not change the
  number of coverage holes in Rips complex of a WSN, we only need to prove that in 
  the process of any HP transformation, there is no new coverage holes created and no two 
  coverage holes merged. If a new coverage hole is created when a vertex $v$ (or edge $e$) is deleted,
  then the boundary cycle of the new coverage hole must be a cycle in $\Gamma_G(v)$ (or $\Gamma_G(e)$),
  which means $\Gamma_G(v)$  (or $\Gamma_G(e)$) is not a simple-connectedness graph. It is contrary to 
  the third condition in HP transformation, so there is no new coverage hole created.
  Furthermore, if two coverage holes are merged when a vertex $v$ (or edge $e$)
  is deleted, then the neighbour graph $\Gamma_G(v)$ ($\Gamma_G(e)$) must not be connected, which
  is contrary to the second condition in HP transformation. So no two coverage holes
  are merged. The situations where two holes would merge after a vertex $v$ deletion are shown in Fig. \ref{figsituation}(a) and (b). Fig. \ref{figsituation}(b) also shows the situation for edge $e$ deletion. Consequently, the number of coverage holes will not be changed
  in the process of any HP transformation.
  \begin{figure}[ht]
   \centering
   \includegraphics[width = 0.35\textwidth]{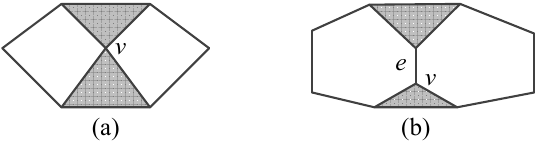}
  \caption{Illustration of merge of two holes when deleting a vertex or edge}
   \label{figsituation}
  \end{figure}
\end{IEEEproof}

\subsection{Weight computation} \label{secweight}
In this component, each node first constructs its simplices to form the Rips complex of the WSN and then computes its weight. For any fence node, its weight is 0.
For any internal node, theoretically the node needs to construct all the simplices which it is part of.
As we consider WSNs in a planar target field, each internal node only needs to construct
all its 1-simplices and 2-simplices and their neighbours. This can also reduce the 
computation complexity. In order to do this, the node 
needs to obtain all its 1- and 2-hop neighbours information. This can be 
achieved by two broadcasts of hello message. In the first one, each node broadcasts 
its id. When it gets all the IDs of its 1-hop neighbours, it continues to 
broadcast a hello message containing the IDs of its 1-hop neighbours. After 
receiving the neighbour list of its neighbours, the node can obtain its $E(v)$,
the set of edges and $T(v)$, the set of 2-simplices.
It can also get the neighbours of each simplex.
For any $e \in E(v)$, let $n(e)$ denote the neighbour set of $e$. For any $t \in T(v)$,
let $n(t)$ denote the neighbour set of $t$. Then the weight of node $v$ can be computed
as in Algorithm \ref{alg1}.

\begin{algorithm} 
\caption{Weight computation (for internal node $v$)}
Begin
\begin{algorithmic}\label{alg1}
\IF {$\exists e \in E(v), n(e)$ is empty}
 \STATE $w_v = 0$
\ELSIF {$\exists t \in T(v), n(t)$ is empty}
 \STATE $w_v = 2$
\ELSE
 \STATE $w_v = 3$
\ENDIF

\end{algorithmic}
END
\end{algorithm}

\subsection{Vertex and edge deletion} \label{secvertex}
In this component, we conduct maximal vertex deletion without changing
the number of coverage holes in the original WSN and also delete some 
special edges if such edges exist. For vertex deletion, we only consider internal nodes, fence nodes will never be deleted.

\subsubsection{Vertex deletion}

As explained in Section \ref{secdefinition}, the larger the weight is, the higher is the probability that the 
node does not lie on the boundary. Meanwhile, if the deletion
of a vertex may create a new coverage hole, it must not be deleted no matter how 
high the weight is. So we have such a rule for vertex deletion. If the weight 
of a vertex is smaller than 3, it should never be deleted. Otherwise, the
vertex continues to check whether it is deletable or not according
to HP transformation. After the verification, the vertex broadcasts a 
message indicating that it can be deleted or not. After receiving
the status of all its neighbours, each deletable vertex continues to check 
whether it should be deleted. The weight of any deletable
vertex must be 3. We assume that the vertex with a lower ID has the priority 
to be deleted first. So each deletable vertex just needs to check whether
its ID is the lowest among all its deletable neighbours. If so, it 
should be deleted. Otherwise, it should not be deleted. Algorithm \ref{alg2} gives
the detailed process for vertex deletion. According to the rule,
two neighbouring vertices will not be deleted simultaneously, so
each vertex can make the decision independently. When a vertex is deleted, it broadcasts
a message to its neighbours. All its neighbours will modify their simplices accordingly and
compute their weights again. The procedure of vertex deletion terminates until
no vertex can be deleted in the Rips complex. Fig.  \ref{procedure}(b)
gives the final result after vertex deletion.

\begin{algorithm} 
\caption{Vertex deletion (for internal node $v$)}
Begin
\begin{algorithmic}\label{alg2}
\IF {$w_v < 3$}
 \STATE node $v$ can not be deleted
\ELSIF {node $v$ is not deletable according to HP transformation}
 \STATE node $v$ can not be deleted
\ELSIF {the ID of node $v$ is the smallest among all its deletable neighbours}
 \STATE node $v$ is deleted
\ENDIF

\end{algorithmic}
END
\end{algorithm}

\subsubsection{Edge deletion}

After vertex deletion, it seems natural to delete all edges which are deletable according to HP transformation. We do not run in this way. On one hand, this may not be useful since deleting edges which are far from coverage holes does not help the detection of boundary cycles of coverage holes. On the other hand, deleting all such edges will increase the complexity of the algorithm. 

Fortunately, it has been proven in \cite{VDM13} that it is possible to make the Rips complex planar by deleting vertices and edges if there is no hole in the Rips complex. For a planar Rips complex, the edge which has at most one neighbour must be on the boundary. This inspires us to check the edges which have at most one neighbour even if it may not be easy to make a Rips complex planar when the Rips complex has holes. It is interesting to find that most edges which have at most one neighbour lie on the boundary of a coverage hole, such as the solid bold edges shown in Fig.  \ref{procedure}(c).  But there exist also some special such edges not lying on the boundary, such as the
dashed bold edges shown in Fig.  \ref{procedure}(c). We try to delete
such special edges. 

\begin{figure}[ht]
 \centering
 \includegraphics[width = 0.3\textwidth]{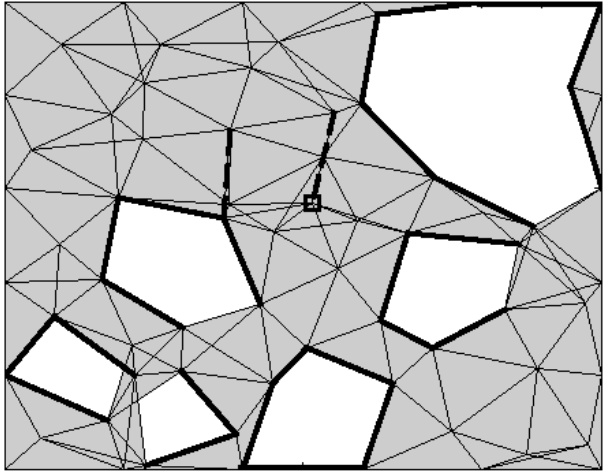}
 \caption{Special boundary edge deletion}
 \label{figedgedeletion}
\end{figure}

We call edges having at most one neighbour as boundary edges, and call boundary edges not lying on the boundary of a coverage hole as special boundary edges. Considering that special boundary edges do not lie on the boundary of a coverage hole, deleting them will not produce new boundary edges. Then we design a rule for deleting special boundary edges. For a boundary edge $[u, v]$ which has only one neighbour $w$\footnote{The boundary edge having no neighbour must be on the boundary.}, if deleting $[u, v]$ will not make $[u, w]$ and $[v, w]$ be boundary edges, then the edge $[u, v]$ can be deleted. We can check that the rule is HP since the neighbouring graph of the edge $[u, v]$ is connected with three vertices $u, v, w$ and without a cycle. It means that deletion of such edges does not change the number of coverage holes in Rips complex. Fig. \ref{figedgedeletion} shows the result of deleting such edges, which is an enlarged version of Fig. \ref{procedure}(c). 
Some edges lying on the boundary may  
also be deleted according to the rule. This is not a big issue, because deletion of such edges just enlarges the current coverage holes. It can be solved in the  
boundary cycles minimization component. 

After edge deletion, some vertices may be deleted again, such as the vertex denoted by a bold square in 
Fig.  \ref{procedure}(c). If such a case happens, we can continue to do vertex deletion until no more vertex or edge can be deleted. Fig.  \ref{procedure}(d) shows the result after edge deletion.

\subsection{Boundary edge detection} \label{secboundaryedge}

As explained in last section, in a planar Rips complex, it is easy to detect all boundary edges by checking whether they have at most one neighbour or not. Then boundary edges are connected sequentially to form boundary cycles. It is thus important to detect boundary edges. In our case, after last step, we can find that some edges lying on the boundary have not been found. This is due to that the Rips complex near coverage holes is not planar. We try to make it planar. 

We consider the nodes having one or more 
boundary edges as boundary nodes and other nodes as non-boundary nodes. First, we need to
delete some edges connecting non-boundary nodes and boundary nodes according to HP
transformation, such as the dashed bold edges shown in Fig. \ref{procedure}(e). After that, 
some new boundary edges may be recognized as shown in Fig. \ref{procedure}(f).
But it is possible that the new found boundary edges cross with non-boundary edges, as the dashed bold edge 
in coverage hole 5 in Fig.  \ref{procedure}(h), or cross with each other, as the dashed bold edges in coverage holes 1 and 2 in Fig.  \ref{procedure}(h). As for the case in coverage hole 5, we can design a similar rule as in last step to delete them. Considering such edges are near the boundary of a coverage hole, deletion of them can make new boundary edges, we design the rule as follows: for a boundary edge $[u, v]$, it has one neighbour $w$, $[u, w]$ and $[v, w]$ are not boundary edges, if the deletion of the edge $[u, v]$ can make
at least one of the two edges $[u, w]$ and $[v, w]$ be boundary edge, then $[u, v]$ can be deleted. Such a rule is HP as explained in last component. As for the crossing boundary edges case, it is illustrated in the next paragraph.

\begin{figure}[ht]
  \centering
  \includegraphics[width=0.4\textwidth]{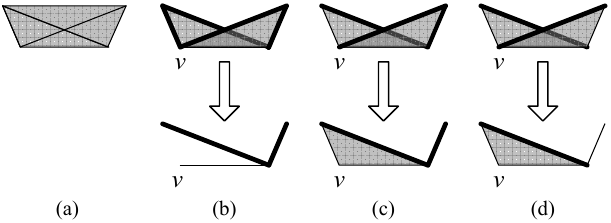}
  \caption{Illustration of crossing boundary edges}
  \label{crossedge}
\end{figure}

Second, we need to delete some edges connecting boundary nodes according
to HP transformation, such as the dashed bold edge in Fig. \ref{procedure}(g). Considering that crossing edges can only exist in 3- or higher dimension simplices, and higer dimension simplices can transfer to several 3-simplices by deleting some edges. We thus only consider 3-simplices, an example is shown in Fig. \ref{crossedge}(a). Deleting one edge of a 3-simplex can produce three possible crossing boundary edges cases\footnote{Here we do not consider the case without crossing boundary edges}, as shown in the top part of Fig. \ref{crossedge} (b) $\sim$ (d),  where bold lines denote boundary edges and other ones denote non-boundary edges. Then we can design corresponding rules to delete some boundary edges. The bottom parts of Fig. \ref{crossedge} (b) $\sim$ (d) give the results after deletion. The rules are also HP as explained in last component.

According to such rules, some boundary edges can be deleted, such as the dashed bold edges in 
Fig. \ref{procedure}(i). From Fig. \ref{procedure}(i), it can be found 
that certain boundary edge is deleted incorrectly. It is not a big issue as explained in 
last component. After deletion of such edges,
new boundary edges can be found as shown in Fig.  \ref{procedure}(j). 

\begin{figure}[ht]
  \centering
  \subfloat[]{\includegraphics[width=0.25\textwidth]{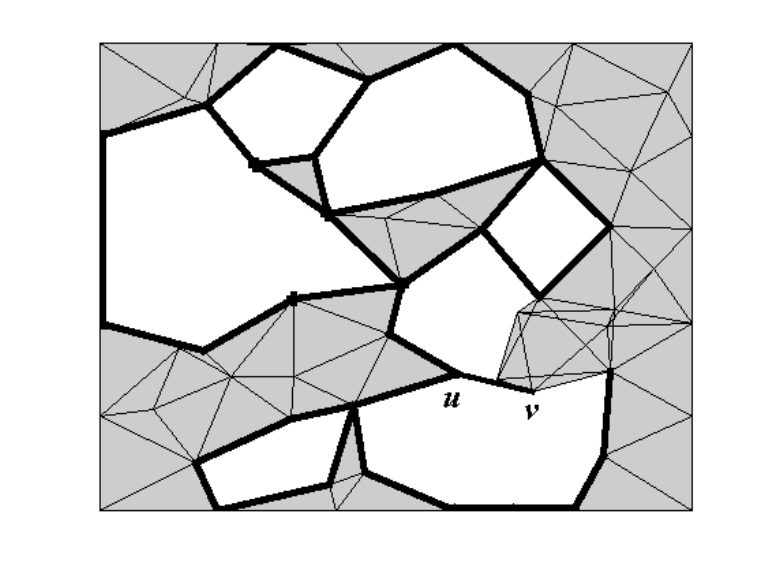}} 
  \hfil
  \subfloat[]{\includegraphics[width=0.15\textwidth]{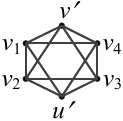}} \\
  \caption{Special case when some boundary edges can not be detected}
  \label{hex}
\end{figure}

In general, after the process above, all boundary edges can be found. But there exists
one special case as in Fig. \ref{hex}(a), where some boundary edges can 
not be detected. This is due to that for a vertex  $v\prime$, as in Fig. \ref{hex}(b), its neighbouring graph is not a simple-connected graph since the cycle formed by $v_1, v_2, v_3, v_4$ can not be triangulated, but the cycle can be triangulated in the graph induced by $v_1, v_2, v_3, v_4, u\prime$. In this case, no vertex or edge is deletable according to HP transformation and no boundary edges can be found since each edge has two neighbours. Such case has no impact on boundary cycles detection, as illustrated in the next component.

\subsection{Coarse boundary cycles discovery}

After boundary edges are detected in the former component, it is easy to discover
the coarse boundary cycles. We just need to randomly choose one node which has
two boundary edges in any boundary cycle. The node initiates the process to find
the coarse boundary cycle by sending a message along one of the boundary edges. Then the boundary
neighbour continues sending the message along its boundary edges.
When the initiating node receives the message coming back along the other boundary edge, it discovers one coarse
boundary cycle. Similarly, all coarse boundary cycles can be found, as shown in Fig.  \ref{procedure}(k). 
If only part of boundary edges are detected after former components, we can not transmit the message  only along boundary edges. The message will be also transmitted along non-boundary edges and some non-boundary nodes will also need to broadcast the message, which increases the complexity of the algorithm.

As for the special case shown in Fig.  \ref{hex}(a), when the node $v$ receives 
a message from its boundary neighbour node $u$, it broadcasts the message to all
its neighbours except $u$. If its neighbour node is a boundary node,
then the message can be sent along the boundary edges. If its neighbour node is
not a boundary node but it has boundary neighbour nodes, then it can send the message
to its boundary neighbour nodes. Else, it will not transmit the message again.
In this way, the message goes along boundary edges most of the time and can return
to the original node initiating the message.

\subsection{Boundary cycles minimization}
It is possible that some coarse boundary cycles found are not 
minimum, so we need to minimize such cycles. This can be achieved by checking
whether there exists a shorter path between any two nodes in the cycle. Since each 
node has its 1- and 2-hop neighbours information, it can locally check the existence
of a shorter path in the cycle in general cases. If there exists, we shorten the cycle and continue to do the same verification until
no such case exists. After that, it is still possible some cycle has not been minimized, 
such as the coverage hole 2 in Fig.  \ref{procedure}(k). So we use the following 2-hop
shrinking process to shorten the cycle. For any four adjacent nodes
in the cycle, say $a, b, c, d$, if there exists one node $x$ which is one common neighbour
of nodes $a, b, c, d$, then the cycle can be shortened by using $x$ to replace nodes $b$ and $c$.

In this way, we can nearly obtain most minimum cycles surrounding coverage holes. In some cases, we can not get the minimum cycles since each node only has its 1- and 2-hop
neighbours information. Even so, the boundary cycles discovered in the algorithm can 
still provide valuable information about coverage holes. %Theoretically, if each node has more information about its $k$-hop neighbours, all cycles can be minimized.

\section{Simulations and performance evaluation}
Performance evaluation of the theoretical bounds obtained in Section IV and 
the algorithm proposed in Section V is presented in this section.

\subsection{Simulation settings}
%We have different simulation settings to evaluate the bounds and the algorithm.

For bounds evaluation, a disk centered at the origin with radius $R_c$ is considered in the simulations.
The probability that the origin is inside a triangular hole is computed.
Sensors are randomly distributed in the disk
according to a Poisson point process with intensity $\lambda$. The sensing
radius $R_s$ of each node is set to be 10 meters and $\gamma$ is
chosen from 2 to 3 with interval of 0.2. So the communication radius
$R_c$ ranges from 20 to 30 meters with interval of 2 meters. $\lambda$ is
selected from 0.001 to 0.020 with interval of 0.001. For each $\gamma$,
$10^7$ simulations are run under each $\lambda$ to check whether the
origin belongs to a triangular hole.

For performance evaluation of the algorithm, we choose a  100 $\times$ 100 m$^2$ square 
area as the target field. The sensing radius $R_s$ of each node is 10 meters. The 
communication radius $R_c$ is set to be 20 meters and so $\gamma = 2$. There are fence 
sensors locating along the edges of the square with 20 meters distance between neighbours. 
Other internal sensors are randomly distributed in the area according to a Poisson point 
process with intensity $\lambda$.

\subsection{Proportion of the area of triangular holes}

\begin{figure}[htbp]
  \centering \subfloat[]{\includegraphics[width=0.35\textwidth]{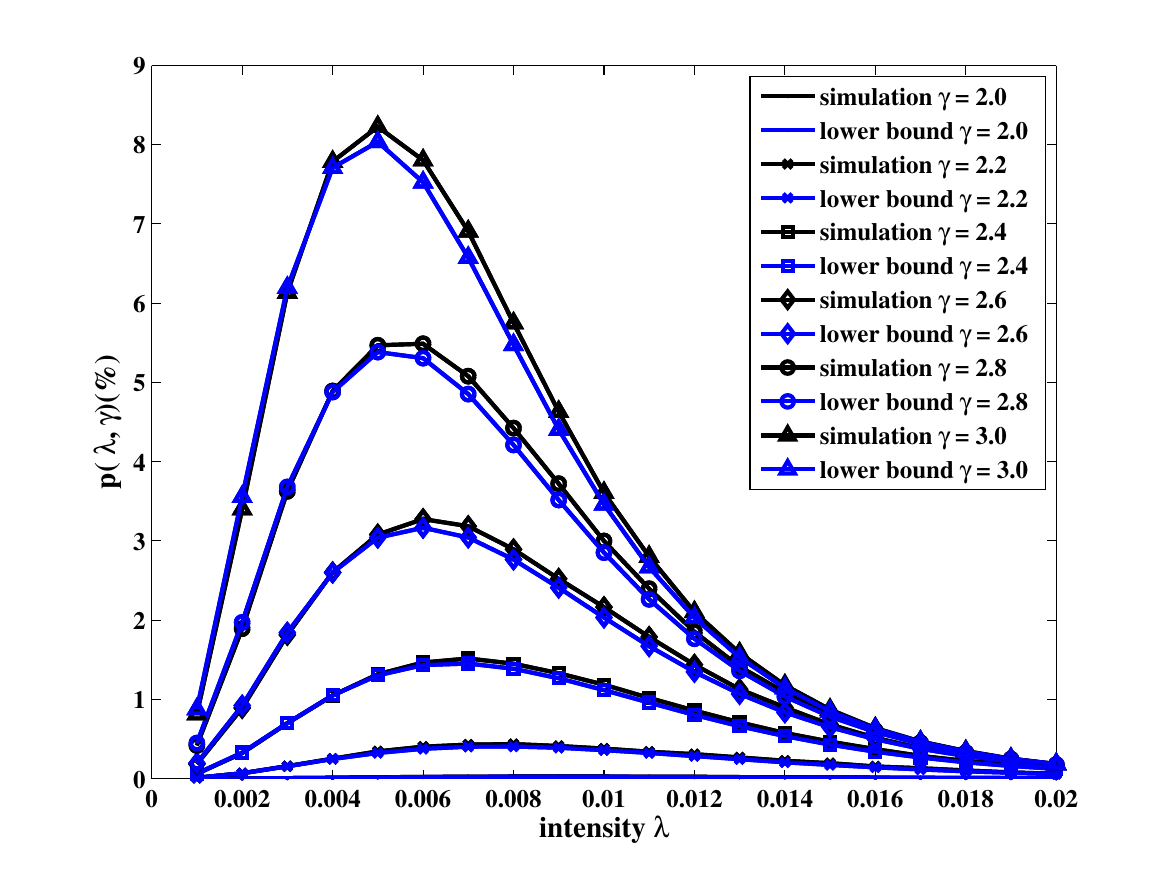}}
  \\
  \subfloat[]{\includegraphics[width=0.35\textwidth]{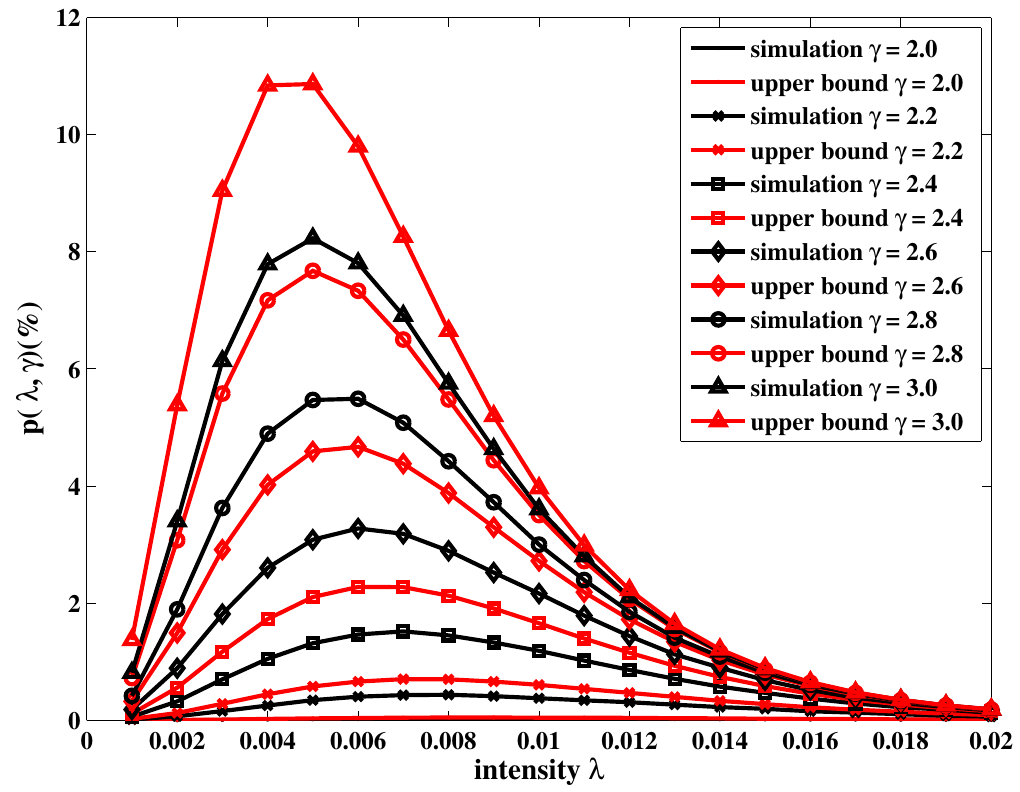}}
  \caption{Proportion of the area of triangular holes (a) simulation results and
    lower bounds ; (b) simulation results and upper bounds }
  \label{prob}
\end{figure}

\begin{figure}[ht]
  \centering
  \includegraphics[width=0.35\textwidth]{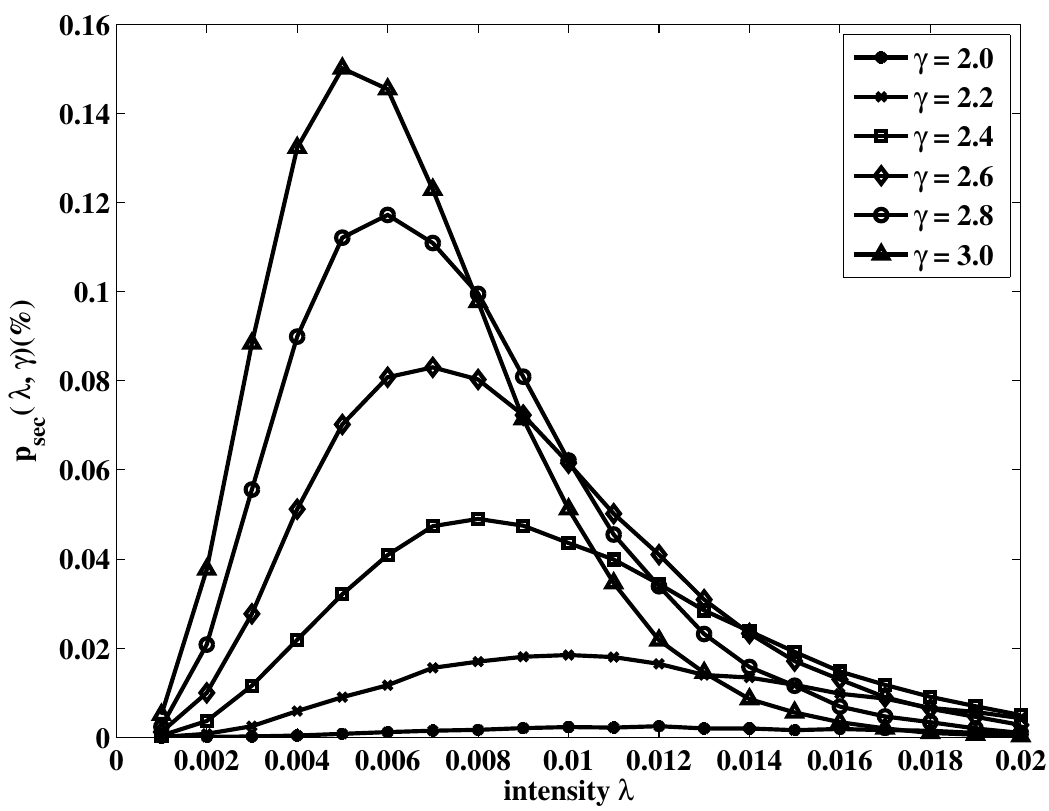}
  \caption{Simulation results for $p_{sec}(\lambda, \gamma)$.}
  \label{probsec}
\end{figure}

The probability $p(\lambda, \gamma)$ obtained by simulations is presented with the lower
and upper bounds in Fig.  \ref{prob}(a) and \ref{prob}(b) respectively. The simulation results for $p_{sec}(\lambda, \gamma)$ are shown in Fig.  \ref{probsec}, which indicate that $p_{sec}(\lambda, \gamma)$ is always smaller than 0.16\% in the simulation settings.

It can be seen from Fig.  \ref{prob} that for any value of $\gamma$, $p(\lambda, \gamma)$ has a maximum at a threshold value $\lambda_{c}$ of the intensity. 
As a matter of fact, for $\lambda \leq \lambda_{c}$, the number of nodes is small. Consequently the probability that the origin belongs to a triangular hole is relatively small too. With the increase of $\lambda$, the connectivity between nodes becomes stronger. As a result, the probability that the origin belongs to a triangular hole increases. However, when the intensity reaches the threshold value, the origin is covered with maximum probability.  $p(\lambda, \gamma)$ decreases for $\lambda \geq \lambda_{c}$. The simulations also show  that $\lambda_{c}$ decreases with the increase of $\gamma$. 

On the other hand, it can be seen from Fig.  \ref{prob}(a) and \ref{prob}(b) that for a fixed intensity $\lambda$, $p(\lambda, \gamma)$ increases with the increases of $\gamma$. That is because when $R_s$ is fixed, the larger $R_c$ is, the higher is the probability of each triangle containing a coverage hole. 

Furthermore, the maximum probability increases quickly with $\gamma$ ranging from 2.0 to 3.0. It is shown that when $\gamma = 2$, the maximum probability from simulation is about 0.03\% and thus it is acceptable to use Rips complex based algorithms to discover coverage holes. While the ratio $\gamma$ is high to a certain extent, it is unacceptable to use connectivity information only to discover coverage holes. 

Finally, it can be found in Fig.  \ref{prob}(a) that the probability obtained by simulation is very well consistent with the lower bound. The maximum difference between them is about 0.5\%. Fig.  \ref{prob}(b) shows that probability obtained by simulation is also
consistent with the upper bound. The maximum difference between them is about 3\%.

In addition, combined with the homology based algorithm, the analytical results can be used for planning of WSNs. For example, assume a WSN is used to monitor a planar target field and 
the ratio $\gamma = 2$, according to the analytical upper bounds, 
we can see that the maximum proportion of the area of triangular
holes under $\gamma = 2$ is about 0.06\%, which can be neglected.
It means that as long as the non-triangular holes can be detected by the 
homology based algorithm and covered by additional nodes, we can say the target field
is covered. But if $\gamma = 3$, it can be seen from the analytical upper bounds that the maximum proportion of the area of triangular holes is about 
11\%, which means that even if the non-triangular holes can be detected and covered, it is still possible that about 11\% of the target field is uncovered. Under this situation, some approaches like increasing the intensity of nodes need to be used in order to make most of the target field covered.

\subsection{Performance of the algorithm}

\subsubsection{Complexity}

The computation complexity of each step in the algorithm is shown in Table \ref{complexity}.
In  weight computation, each node needs to construct its 0-, 1- and 2-simplices. For construction of 0- and 1-simplices, each node only needs to know its 1-hop neighbour information, which can be done by a broadcast as explained in Section \ref{secweight}. For 2-simplices construction, each node needs to obtain its neighbours' 1-hop neighbour information, which is achieved by another broadcast. Then the node continues to determine whether it can form a 2-simplex with any two of its neighbours by checking whether they have a common neighbour. The node needs to check $n(n-1)/2$ times, where $n$ is the number of its 1-hop neighbours. Then the node computes its weight by checking all its 2-simplices of which the maximum number is $n(n-1)/2$.
The computation complexity of this component is thus $O(n^2)$.

In vertex deletion part, each node needs to check whether it is deletable or not according
to HP transformation. This can be done by checking all its cycles in its neighbouring graph.
It can build a spanning tree in its neighbouring graph and check all fundamental cycles
in the spanning tree. There are $E-n+1$ fundamental cycles, where $E$ is the number 
of edges in its neighbouring graph, so the worst case computation complexity is $O(n^2)$.
Since the node needs to recompute its weight and recheck whether it is deletable when any
one of its neighbour is deleted, so the total worst case computation complexity is $O(n^3)$.
As for edge deletion, the node only needs to check all its 1-simplices to see whether there exists
the case illustrated in Section  \ref{secvertex}, so the complexity is $O(n)$. The complexity of this component is thus $O(n^3)$.

In the boundary edge detection component, nodes need to check whether some of their edges can be deleted or not according to HP transformation. For each edge, the worst case computation complexity is $O(n^2)$ as explained in last paragraph, the total worst case computation complexity is thus $O(n^3)$ since there are maximum $n$ edges. The actual complexity is much less than that since for one edge, there are usually very few nodes in its neighbouring graph. In addition, the boundary nodes need to check whether there exist special cases
as illustrated in Section \ref{secboundaryedge}. The node needs to check all its 2-simplices, which is of complexity $O(n^2)$ since there are maximum $n(n-1)/2$ 2-simplices. So the complexity of this component is $O(n^3)$.

As for the final two components, each node only needs to 
broadcast some messages and do some local computations, the complexity is $O(1)$.
So the total worst case computation complexity for our algorithm is $O(n^3)$.

\begin{table}
  \caption{complexity of each step in the algorithm}
  \label{complexity}
  \centering
  \begin{tabular}{|c|c|}
  	 \hline
     Step & Complexity \\
     \hline
     Weight computation & $O(n^2)$ \\
     \hline
     Vertex and edge deletion &  $O(n^3)$ \\
     \hline 
     Boundary edge detection &  $O(n^3)$\\
     \hline
     Coarse boundary cycles discovery &  $O(1)$\\
     \hline 
     Boundary cycles minimization &  $O(1)$ \\
     \hline
  \end{tabular}
\end{table}

\subsubsection{Comparison with other algorithms}

In order to evaluate the performance of our proposed homology based algorithm (denoted as HBA), 
we compare it with
the location based algorithm (denoted as LBA) proposed in \cite{TT06}.
Since location based algorithm can discover both triangular and non-triangular coverage holes, but 
our algorithm can only detect non-triangular coverage holes, we do
not consider those triangular coverage holes in the comparison. It is 
possible that there exist shorter paths in boundary cycles found by LBA, we first
shrink them using 1-hop neighbour information of boundary nodes. After that,
we compare those boundary cycles with what our algorithm finds. For
some coverage holes, the minimum boundary cycles may not be unique, two boundary
cycles are considered to surround the same coverage hole if one cycle can be
converted to another one by using only 1-hop neighbours information. 
We emphasize that only 1-hop neighbours information can be used in the comparison
in order to evaluate the accuracy of boundary cycles found by our algorithm. For example,
if one cycle $c_1$ found by our algorithm can not be converted to another cycle $c_2$ found by LBA using only 1-hop neighbours
information but can be converted by using 2-hop neighbours information, we consider
the cycle $c_1$ is not accurate and the corresponding coverage hole is not found.

\begin{figure}[ht]
  \centering
  \includegraphics[width=0.35\textwidth]{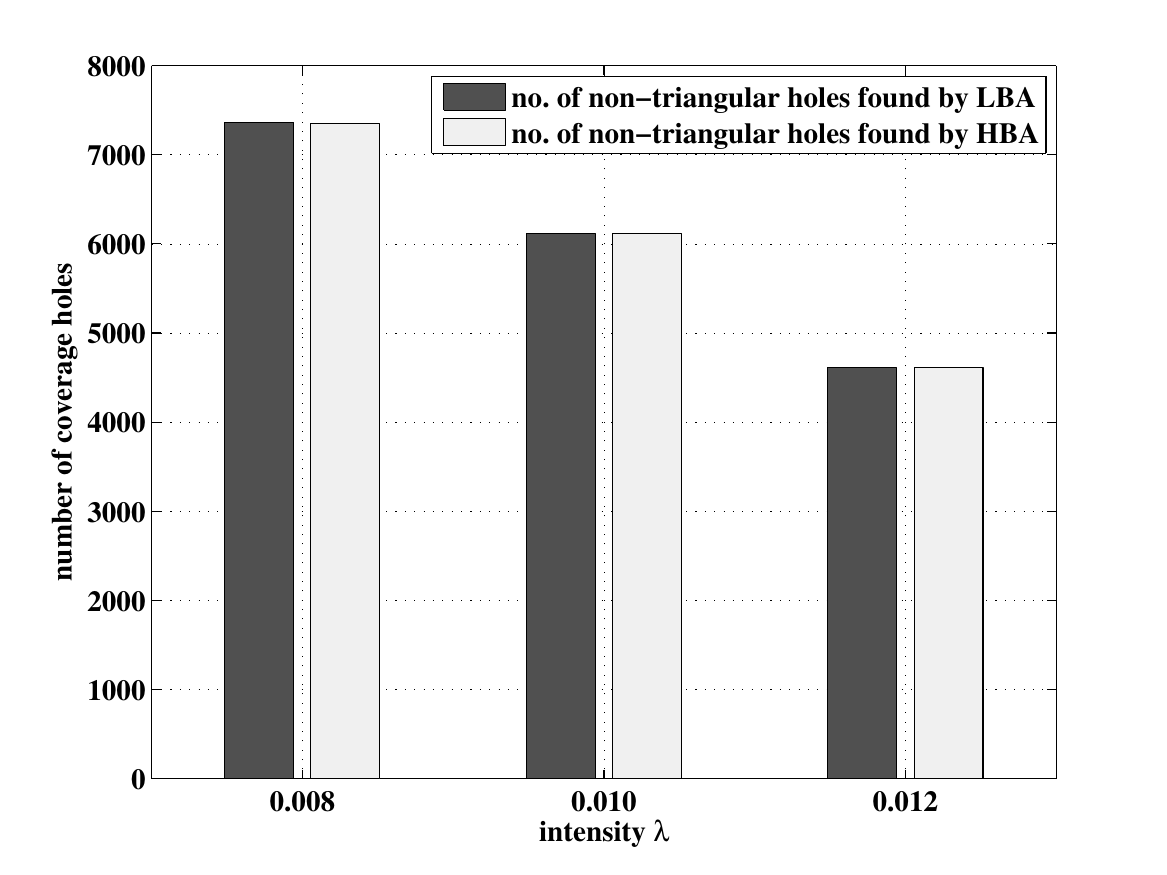}
  \caption{Performance evaluation of the algorithm}
  \label{perf}
\end{figure}

Based on the method presented above, we set $\lambda$ to be 0.008, 0.010 and 0.012 
to represent sparse, moderate and dense WSNs respectively. For each intensity, 1000
simulations are performed. Simulation results show that when $\lambda$ is 0.008,
there are nine times among the 1000 times when our algorithm can not find all non-triangular 
coverage holes. In each of the nine times, only one coverage hole is missed. 
There are 7363 non-triangular holes in total and 7354 ones found by our algorithm.
When $\lambda$ is 0.010 and 0.012, only one time among the 1000 times when our algorithm 
can not find all coverage holes. And in that time, only one coverage hole is missed.
When $\lambda$ is 0.010, there are 6114 non-triangular holes in total and 6113 ones
found by our algorithm. When $\lambda$ is 0.012, there are 4613 non-triangular holes in 
total, of which 4612 ones are found. The results are shown in Fig.  \ref{perf}.
All these results show that our algorithm can find about 99\% coverage holes in 
about 99\% cases.

\section{Conclusions}

In this paper, we adopt two types of simplicial complex called $\check{\textrm{C}}$ech 
complex and Rips complex to capture coverage holes of a WSN. The relationship between 
$\check{\textrm{C}}$ech complex and Rips complex in terms of coverage hole is first 
analysed under different ratios between communication radius and sensing radius of a 
sensor. Based on that, we define two types of coverage holes: triangular and 
non-triangular hole. For triangular holes, both the lower and upper bounds on the 
proportion of the area of triangular holes in a WSN are derived. Such proportion is 
related to the ratio  between communication radius and sensing radius
of each sensor. When the ratio is no larger than $\sqrt{3}$, there is no triangular hole. 
When  the ratio is between $\sqrt{3}$ and 2, both the theoretical analysis and simulation
results show that the proportion is lower than 0.06\% under any intensity. It means that 
the triangular holes can nearly be neglected. When the ratio is larger than 2, the 
proportion of the area of triangular holes increases with $\gamma$. It becomes unacceptable 
for $\gamma$ larger than a threshold. In that case triangular holes can not be neglected 
any more. For non-triangular holes, a homology-based algorithm is proposed to detect 
them. Simulation results show  that the algorithm can detect 99\% such holes.

\appendices
\section{Proof of Theorem \ref{trihole2}} \label{app1}
\begin{IEEEproof}:
We first prove the lower bound. It can be obtained from (\ref{eq0}) that
\begin{equation*} 
  \begin{split}
    p(\lambda, \gamma) > \mathrm{P}\{\bigcup_{\{n_1, n_2\} \subseteq \Phi
      \backslash \{\tau_0(\Phi)\}} T(\tau_0, n_1, n_2)\} 
  \end{split}
\end{equation*}
So for the lower bound, we only consider the case that the closest node $\tau_0$ can contribute to a triangle which bounds a triangular hole containing the origin. 

Using polar coordinates, we assume the closest node $\tau_0$ lies on $(d_0, \pi)$. It is well known that the distance $d_0$ is a random variable with distribution
\begin{equation} \label{eq1}
  F_{d_0}(r_0) = \mathrm{P} \{ d_0 \leq r_0\} = 1 - e^{- \lambda \pi r_0^2}
\end{equation}

Therefore the probability of the first case can be given as
\begin{equation} \label{eq2}
  \begin{split}
    & \mathrm{P}\{\bigcup_{\{n_1, n_2\} \subseteq \Phi \backslash \{\tau_0(\Phi)\}} T(\tau_0, n_1, n_2)\} \\
    = & \int \mathrm{P}\{\bigcup_{\{n_1, n_2\} \subseteq
      \Phi_{r_0}^\prime} T((r_0, \pi), n_1, n_2)\} F_{d_0}(dr_0)
  \end{split}
\end{equation}
\noindent where $\Phi_{r_0}^\prime$ is the restriction of $\Phi$ in $B(O, R_c)
\backslash B(O, r_0)$. 

Once the node $\tau_0$ is determined, a second node $\tau_1$ must lie in the shadow 
region $A^+$ shown in Fig.  \ref{case2lower2} and a third node $\tau_2$ must lie in the region $S^-$ shown in Fig.  \ref{case2lower2}, as illustrated in Section \ref{seccase2}. The node $\tau_1 = (d_1, \theta_1)$ is assumed to have the smallest
polar angle in $A^+$, which means that there should be no nodes with a polar angle
less than $\theta_1$ in $A^+$, that is to say no nodes are in the region
\begin{equation*}
  S^+(\tau_0, \tau_1) =  S^+(d_0, \theta_1) = A^+ \bigcap H^+(\theta_1)
\end{equation*}

Since the intensity measure of the PPP in polar
coordinates is $\lambda r dr d\theta$, the density $F_{\tau_1}$ of
$\tau_1$ can be given as
\begin{equation} \label{eq3}
  F_{\tau_1} (dr_1, d\theta_1) = \lambda r_1 e^{-\lambda |S^+(d_0, \theta_1)|} dr_1 d\theta_1
\end{equation}

The integration domain $D(d_0)$ with respect to parameters $(d_1,
\theta_1)$ can be easily obtained. From the construction of the region
$A^+$, we can get $\alpha_0 = 2\arccos(R_c/(2d_0))$ and 
$\alpha_1 = 2\arcsin(R_c/(2d_0)) - 2\arccos(R_c/(2d_0))$. So
$\alpha_0 \leq \theta_1 \leq \alpha_1$ and $d_0 < d_1 \leq
R_1(d_0, \theta_1)$, where
\begin{equation*}
  \begin{split}
  R_1(d_0, \theta_1) = &\min(\sqrt{R_c^2 - d_0^2\sin^2 \theta_1} - d_0 \cos \theta_1,\\
       & \sqrt{R_c^2 - d_0^2\sin^2 (\theta_1+\alpha_0)} + d_0 \cos (\theta_1+\alpha_0))
  \end{split} 
\end{equation*}

Furthermore, the area $|S^+(r_0, \theta_1)|$ can be expressed as
\begin{equation*}
    |S^+(r_0, \theta_1)| = \int_{\alpha_0} ^{\theta_1} \int_{r_0}^{R_1(r_0, \theta)} r dr d\theta
\end{equation*}

As illustrated in Section \ref{seccase2}, assume only $\tau_0, \tau_1$ 
and nodes in $S^-(\tau_0, \tau_1)$ can
contribute to the triangle which bounds a triangular hole containing the
origin, we can get a lower bound of the probability that the origin is inside a  triangular hole. Based on the assumption, we have
\begin{equation} \label{eq4}
  \begin{split}
    &\mathrm{P}\{\bigcup_{\{n_1, n_2\} \subseteq \Phi_{r_0}^\prime} T((r_0, \pi), n_1, n_2)\}  \\
    & > \mathrm{P}\{\bigcup_{n_2 \subseteq \Phi_{r_0}^\prime \bigcap S^-(\tau_0, \tau_1) } T((r_0, \pi), \tau_1, n_2)\}\\
    & = \iint_{D(r_0)} \mathrm{P} \{ \bigcup_{n_2 \subseteq \Phi_{r_0}^\prime \bigcap \atop S^-(r_0, r_1, \theta_1)} T((r_0, \pi), (r_1, \theta_1), n_2) \}  F_{\tau_1} (dr_1, d\theta_1) \\
    & = \iint_{D(r_0)} \mathrm{P} \{\Phi_{r_0}^\prime (S^-(r_0,
    r_1, \theta_1)) > 0 \} F_{\tau_1} (dr_1, d\theta_1) \\
    & = \iint_{D(r_0)} (1-e^{-\lambda|S^-(r_0, r_1, \theta_1|}) F_{\tau_1} (dr_1, d\theta_1)
  \end{split}
\end{equation}
\noindent where $|S^-(r_0, r_1, \theta_1|$ can be expressed as 
\begin{equation*} \label{eqareasminus}
 |S^-(r_0, r_1, \theta_1)| = \int_{\theta_{2l}}^{-\alpha_0} \int_{r_0}^{R_2(r_0, r_1, \theta_1, \theta_2)}r dr d\theta_2
\end{equation*}
and 
\begin{align*}
	& \theta_{2l} = \theta_1 - \arccos \frac{\cos (R_c/R) - \cos \theta_1 \cos \theta_0}{\sin \theta_1 \sin \theta_0} \\
	& R_2(r_0, r_1, \theta_1, \theta_2) = \min(\sqrt{R_c^2 - r_0^2\sin^2 \theta_2} - r_0 \cos \theta_2, \\
	& \qquad \sqrt{R_c^2 - r_1^2\sin^2 (\theta_2-\theta_1)} + r_1 \cos (\theta_2-\theta_1) )
\end{align*} 

Therefore, from (\ref{eq1}), (\ref{eq2}), (\ref{eq3}) and (\ref{eq4}), 
the lower bound shown in (\ref{lower1}) can be derived.

As for the upper bound, replace $|S^-(r_0, r_1, \theta_1)|$ by 
$|S^-(r_0, r_0, \theta_1)|$, we can get the upper bound as illustrated in Section \ref{seccase2}.

\end{IEEEproof}

%\section{}


\ifCLASSOPTIONcaptionsoff
  \newpage
\fi



%\bibliographystyle{IEEEtran}
\bibliographystyle{IEEEtran}
\bibliography{IEEEabrv,TON}

\end{document}